\documentclass[review]{elsarticle}

\usepackage{lineno,hyperref}
\modulolinenumbers[1]

\journal{Journal of Nuclear Instruments and Methods in Physics Research A}

\usepackage{xcolor}

\begin{document}

\begin{frontmatter}

\title{The JSNS$^2$ Detector}

\author[15]{S.~Ajimura}
\author[14]{M.~Botran}
\author[4]{J.~H.~Choi}
\author[3]{J.~W.~Choi}
\author[18]{M.~K.~Cheoun}
\author[21]{T.~Dodo}
\author[21]{H.~Furuta}
\author[12]{J.~Goh}
\author[7]{K.~Haga}
\author[7]{M.~Harada}
\author[7,22]{S.~Hasegawa}
\author[21]{Y.~Hino}
\author[15]{T.~Hiraiwa}
\author[16]{H.~I.~Jang}
\author[6]{J.~S.~Jang}
\author[3]{M.~C.~Jang}
\author[19]{H.~Jeon}
\author[19]{S.~Jeon}
\author[3]{K.~K.~Joo}
\author[14]{J.~R.~Jordan}
\author[19]{D.~E.~Jung}
\author[17]{S.~K.~Kang}
\author[7]{Y.~Kasugai}
\author[10]{T.~Kawasaki}
\author[8]{E.~J.~Kim}
\author[3]{J.~Y.~Kim}
\author[19]{S.~B.~Kim\corref{sbk}}
\cortext[sbk]{Corresponding author}
\ead{sbk@snu.ac.kr}
\author[13]{W.~Kim}
\author[7]{H.~Kinoshita}
\author[10]{T.~Konno}
\author[9]{D.~H.~Lee}
\author[12]{S.~Lee}
\author[3]{I.~T.~Lim}
\author[14]{E.~Marzec}
\author[9]{T.~Maruyama}
\author[7]{S.~Masuda}
\author[7]{S.~Meigo}
\author[9]{S.~Monjushiro}
\author[3]{D.~H.~Moon}
\author[15]{T.~Nakano}
\author[11]{M.~Niiyama}
\author[9]{K.~Nishikawa\fnref{Nish}}
\fntext[Nish]{Deceased}
\author[15]{M.~Nomachi}
\author[4]{M.~Y.~Pac}
\author[9]{J. S. Park}
\author[20]{S.~J.~M.~Peeters}
\author[5]{H.~Ray}
\author[19]{G.~Roellinghoff}
\author[19,23]{C.~Rott}
\author[7]{K.~Sakai}
\author[7]{S.~Sakamoto}
\author[15]{T.~Shima}
\author[3]{C.~D.~Shin}
\author[14]{J.~Spitz}
\author[1]{I.~Stancu}
\author[15]{Y.~Sugaya}
\author[21]{F.~Suekane}
\author[7]{K.~Suzuya}
\author[9]{M.~Taira}
\author[21]{R.~Ujiie}
\author[7]{Y.~Yamaguchi}
\author[2]{M.~Yeh}
\author[4]{I.~S.~Yeo}
\author[12]{C.~Yoo}
\author[19]{I.~Yu}
\author[3]{A.~Zohaib}

\address[1]{Department of Physics and Astronomy, University of Alabama, Tuscaloosa, AL, 35487, USA}
\address[2]{Brookhaven National Laboratory, Upton, NY, 11973--5000, USA}
\address[3]{Department of Physics, Chonnam National University, Gwangju, 61186, KOREA}
\address[4]{Laboratory for High Energy Physics, Dongshin University, Chonnam 58245, KOREA}
\address[5]{Department of Physics, University of Florida, Gainesville, FL, 32611, USA}
\address[6]{Department of Physics and Optical Science, Gwangju Institute of Science and Technology, Gwangju, 61005, KOREA}
\address[7]{J-PARC Center, JAEA, Tokai, Naka Ibaraki 319--1195, JAPAN}
\address[22]{Advanced Science Research Center, JAEA, Tokai, Naka Ibaraki 319--1195, JAPAN}
\address[8]{Division of Science Education, Jeonbuk National University, Jeonju, 54896, KOREA}
\address[9]{High Energy Accelerator Research Organization (KEK), Tsukuba, Ibaraki 305--0801, JAPAN}
\address[10]{Department of Physics, Kitasato University, Sagamihara, Kanagawa 252--0373, JAPAN}
\address[11]{Department of Physics, Kyoto Sangyo University, Kyoto 603--8555, JAPAN}
\address[12]{Department of Physics, Kyung Hee University, Seoul 02447, Korea}
\address[13]{Department of Physics, Kyungpook National University, Daegu 41566, KOREA}
\address[14]{Department of Physics, University of Michigan, Ann Arbor, MI, 48109, USA}
\address[15]{Research Center for Nuclear Physics, Osaka University, Osaka 565--0871, JAPAN}
\address[16]{Department of Fire Safety, Seoyeong University, Gwangju 61268, KOREA}
\address[17]{School of Liberal Arts, Seoul National University of Science and Technology, Seoul, 139--743, KOREA}
\address[18]{Department of Physics, Soongsil University, Seoul 06978, KOREA}
\address[19]{Department of Physics, Sungkyunkwan University, Suwon 16419, KOREA}
\address[20]{Department of Physics and Astronomy, University of Sussex, BN1 9QH, Brighton, UK}
\address[21]{Research Center for Neutrino Science, Tohoku University, Sendai, Miyagi 980--8577, JAPAN}
\address[23]{Department of Physics and Astronomy, University of Utah, Salt Lake City, UT, 84112, USA}


\begin{abstract}

The JSNS$^2$ (J-PARC Sterile Neutrino Search at J-PARC Spallation Neutron Source)
experiment aims to search for oscillations involving a sterile neutrino in the eV$^2$ mass-splitting range.
The experiment will search for the appearance of electron antineutrinos oscillated from muon antineutrinos.
The electron antineutrinos are detected via the inverse beta decay process using a liquid scintillator
detector.
A 1\,MW beam of 3\,GeV protons incident on a spallation neutron target produces
an intense and pulsed neutrino source from pion, muon, and kaon decay at rest.
The JSNS$^2$ detector is located 24\,m away from the neutrino source and began operation from June 2020.
The detector contains 17 tonnes of gadolinium (Gd) loaded liquid scintillator (LS) in an acrylic vessel, as a neutrino target. It is surrounded by  31 tonnes of unloaded LS in a stainless steel tank.
Optical photons produced in LS are viewed by 120 R7081 Hamamatsu 10-inch
Photomultiplier Tubes (PMTs).
In this paper, we describe the JSNS$^2$ detector design, construction, and operation.

\end{abstract}


\end{frontmatter}


\section{Introduction and Overview}

The existence of sterile neutrinos remains an open question.
Some experimental results may not be explained by the three active flavor neutrino framework and suggest the existence of sterile a neutrino with a mass-squared splitting around 1~eV$^2$~\cite{LSND,GALLAX,SAGE,RAA,MiniBooNE}.
The JSNS$^2$ experiment will search for sterile neutrino oscillations at the J-PARC Materials and Life science experimental Facility (MLF).
With a 3\,GeV proton beam from the Rapid Cycling Synchrotron (RCS) and a spallation neutron target, an intense and 25\,Hz pulsed neutrino source is available from pion, muon, and kaon decays at rest~\cite{proposal-TDR}.
The experiment will look for appearance of $\overline{\nu}_e$ oscillated from $\overline{\nu}_\mu$ that has a well-known spectrum and a 52.8\,MeV end point energy. The $\overline{\nu}_\mu$ to $\overline{\nu}_e$ oscillation results in $\overline{\nu}_e$ spectral modulation. The produced $\overline{\nu}_e$ is detected by the inverse beta decay (IBD) interaction $\overline{\nu}_e + p \rightarrow e^+ + n$, followed by $\sim$8\,MeV gammas from neutron capture on Gd. The $\overline{\nu}_e$ energy is reconstructed by the positron visible energy plus 0.8 MeV. In particular, the delayed signal identification by neutron capture on Gd allows for a significant reduction of ambient $\gamma$-ray backgrounds below 3\,MeV.

A coincidence requirement of 100\,$\mu$s between prompt and delayed events further reduces the background event rate substantially.
The requirements of the time coincidence and their spatial correlation reduces the accidental background by a factor of approximately $10^{4}$. By taking advantage of the short pulse width and low frequency of the RCS proton beam, a timing gate requirement of a prompt candidate event between 1 and 10\,$\mu$s from the beam starting time eliminates most of the neutrinos from pion and kaon decays and beam-induced fast neutrons, and also reduces the cosmic-ray induced background by $\sim$$10^{5}$.

\begin{figure}[hbt]
\begin{center}
\includegraphics[width=0.8\textwidth]{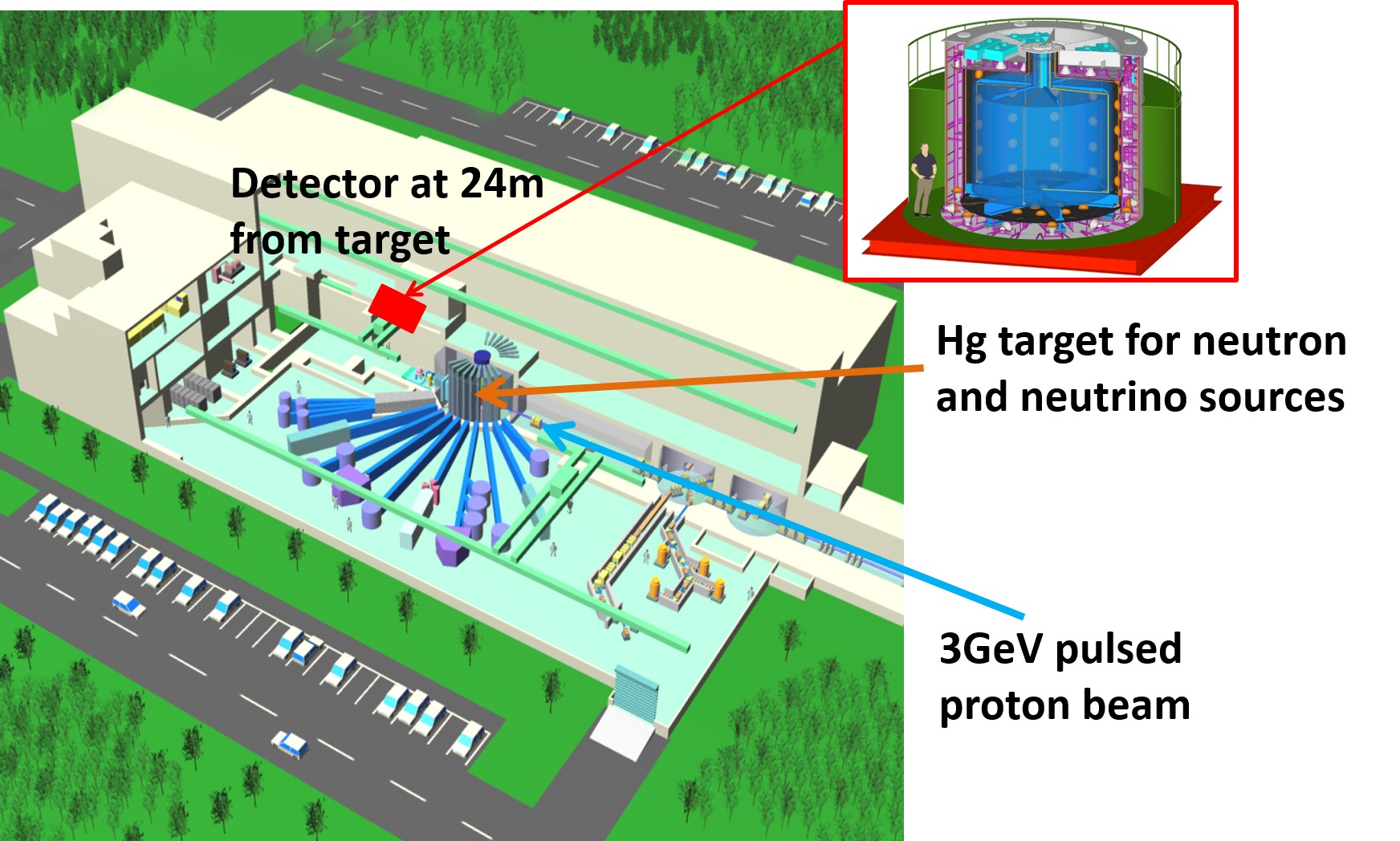}
\caption{Layout of JSNS$^2$ experiment. The detector is located on the third floor of the MLF at a baseline distance of 24\,m from the neutrino source.}
\label{fig:jsns2-layout}
\end{center}
\end{figure}

The detector is located on the third floor of the MLF at a baseline distance of 24\,m from the neutrino source. Figure~\ref{fig:jsns2-layout} shows the overall layout of the JSNS$^2$ experiment.
Data collected with the JSNS$^2$ experiment will be sensitive to eV-scale sterile neutrinos and can be used to perform an ultimate and direct test of the LSND anomaly.

\begin{figure}[t]
 \centering
 \includegraphics[width=0.85 \textwidth]{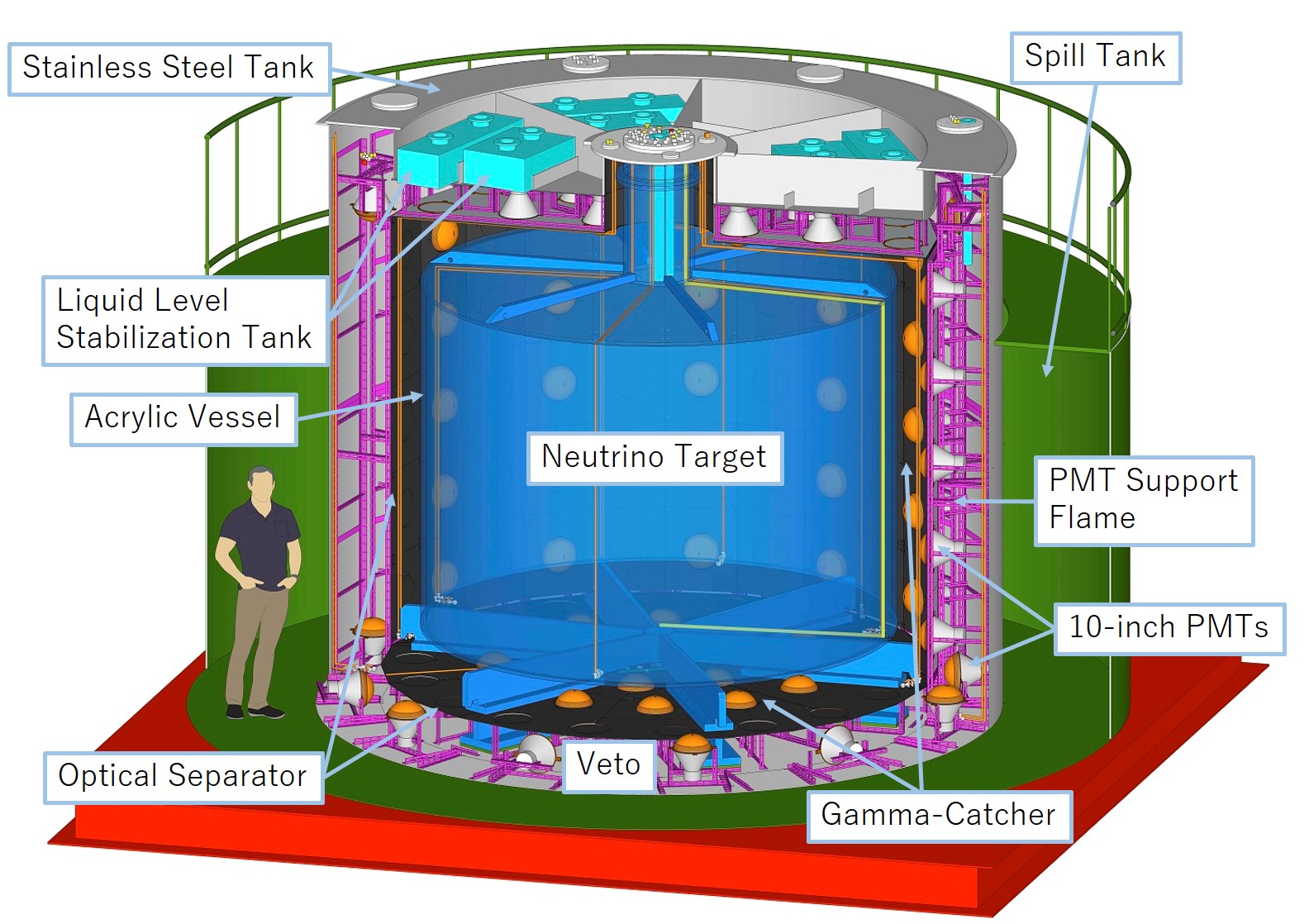}
\caption{\setlength{\baselineskip}{4mm}
  Schematic view of the JSNS$^2$ detector.
}
 \label{Fig:DS}
\end{figure}

The JSNS$^2$ detector consists of three cylindrical layers including an innermost neutrino target, an intermediate gamma-catcher, and an outermost veto. 
The neutrino target is made of 17 tonnes of Gd loaded LS (Gd-LS) stored in an acrylic vessel, 
3.2\,m ($D$) $\times$ 2.5\,m ($H$).
The gamma-catcher and veto are filled with 31 tonnes of Gd unloaded LS in a stainless steel vessel, 4.6\,m ($D$) $\times$ 4.0\,m ($H$), and optically separated. The gamma-catcher is 25\,cm wide while the veto has a thickness of 45\,cm and 25\,cm for the barrel side and for the top and bottom sides, respectively. Optical photons produced via scintillation in the target and gamma-catcher are viewed by an array of 96 10-inch PMTs. Scintillation light in the veto can be detected by an array of 24 10-inch PMTs.
A schematic view of the detector is shown in Fig~\ref{Fig:DS}.

The third floor of the MLF is normally used as a maintenance area for the spallation neutron target and its ancillary systems. During the maintenance period, typically from July to September, the JSNS$^2$ detector must be removed from this area
in order to avoid possible interference with necessary maintenance activity. Therefore, the detector has been designed for a convenient relocation, including liquid filling and extraction.
The detector is assembled and stored in a detector storage building at J-PARC. The detector is transported to the first floor of the MLF using a low-bed trailer. There it is filled with LS and Gd-LS. The 70-tonne detector is lifted to the third floor by a crane and placed on top of a shielding layer.
Lead blocks and iron plates are installed underneath the detector in order to shield against gamma rays produced in the concrete floor by neutron captures coming from the beam target.
During the maintenance period, the detector and shielding materials are removed from the third floor. The detector is lifted back to the first floor where the liquid is extracted into storage tanks. Next, it is transported back to the detector storage building.

The detector construction began in 2017 and was completed in 2020. The first data-taking period in a commissioning mode lasted for ten days in June 2020.  Data taking was resumed with a 0.6 MW proton beam in January 2021. The beam power was increased to 0.7 MW in April 2021 and is expected to reach 1 MW in the near future. In this paper, we describe the design, assembly, and operation of the JSNS$^2$ detector.

\section{Detector Design}

As mentioned in the previous section, the detector consists of three cylindrical layers of neutrino target, gamma-catcher and veto.
This section presents a detailed description of the detector design.

\subsection{Overall Design}
\indent

A schematic view of the JSNS$^2$ detector is shown in Fig~\ref{Fig:DS}.
From inner to outer detector, the experiment is structured as follows: an acrylic vessel holding the target liquid, an array of 10-inch PMTs,
an optical separation wall of black boards, a veto layer,
a stainless steel tank holding the gamma-catcher and veto LS, and a spill tank.
Each of them is described in more detail in the following sections.

\begin{figure}[h]
 \centering
 \includegraphics[width=0.65 \textwidth]{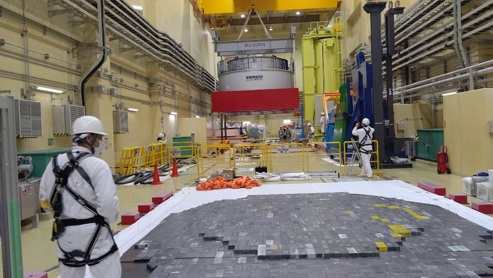}
\caption{\setlength{\baselineskip}{4mm}
  Gamma-ray shielding layers consisting of iron plates and lead blocks.
The 70 tonne detector with  Gd-LS and LS was transported by a crane and
installed on the shielding layers.
}
 \label{Fig:IL}
\end{figure}

Figure~\ref{Fig:IL} shows a picture of a gamma-ray shielding layer underneath the detector.
The gamma-rays are produced in beam neutron capture reactions inside the concrete floor.
Iron plates are assembled into three layers of approximately 6000 $\times$ 9000 $\times$ 66 mm$^3$. Lead blocks are put together to form two disks each with thickness of 50\,mm: a disk of 5\,m in diameter in the lower layer and a disk of 3.7\,m in diameter in the upper layer. This shielding structure reduces the gamma-ray background by a factor of more than 100\,in the 7 to 12\,MeV region.

\subsection{Stainless Steel Tank}
\indent

\begin{figure}[h]
 \centering
 \includegraphics[width=0.75 \textwidth]{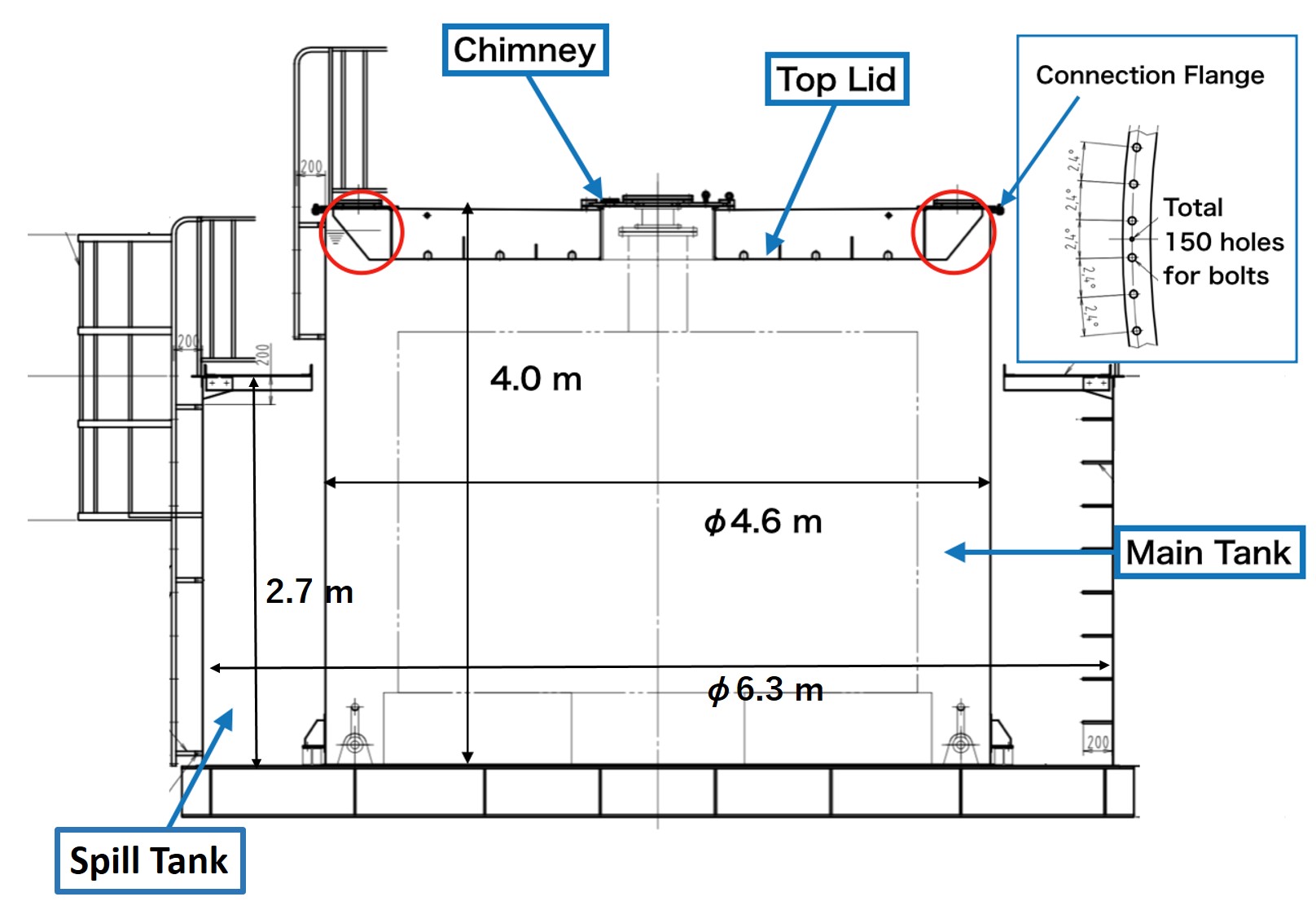}
\caption{\setlength{\baselineskip}{4mm}
  A schematic drawing of the stainless tank and the spill tank.
  The red circles indicate an LS buffer space of a ring shape in order to minimize changes of the liquid level due to thermal expansion of the LS.
}
 \label{Fig:SStankDrawings}
\end{figure}

The 5 mm thick stainless steel tank was constructed by the Morimatsu company in the period 2017 to 2018.
Water leak tests were done to confirm that the tank had no leaks.
Figure~\ref{Fig:SStankDrawings} shows a schematic drawing of the
stainless steel tank and the spill tank.
The detector size is limited by the size of the MLF entrance, which is 7\,m wide and 5\,m high.

The detector is filled with liquid above the top-lid level in order to minimize LS sloshing during detector relocation by a crane.
The detector also has a compact liquid level stabilization system using a reverse siphon system with a minimal pipe length.
As indicated by the red circles in Fig.~\ref{Fig:SStankDrawings},  a ring-shaped LS buffer space is included to provide a large liquid surface and minimize the change of liquid level due to the thermal expansion of LS.
More details of the stainless steel tank are given in Ref.~\cite{CITE:SStank}.

\subsection{Spill Tank}
\indent

To adhere to the local safety regulations, the stainless steel tank is surrounded by a spill tank.
The tank's diameter is 6.3\,m and the inner height is 2.7\,m.
If all the LS leaked out of the stainless steel tank, the liquid level in the spill tank would still be around 1\,m below the top of the tank.
A more detailed description of the spill tank can be found in Ref.~\cite{CITE:SStank}.

\subsection{Moving Equipment}
\indent

\begin{figure}[!h]
 \centering
 \includegraphics[width=0.60 \textwidth]{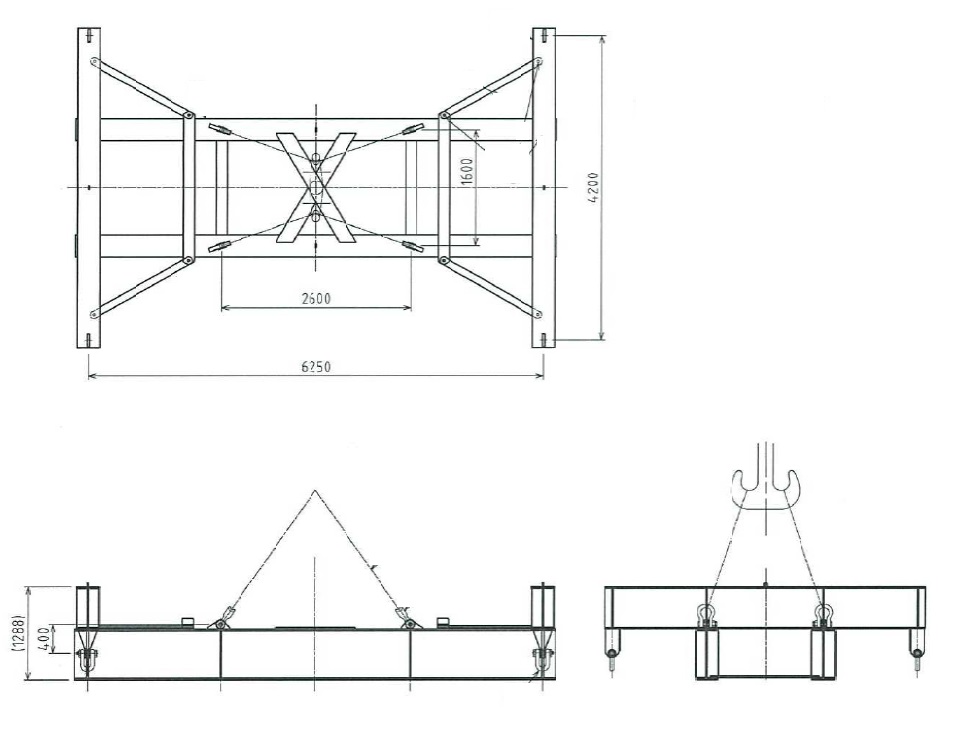}
\caption{\setlength{\baselineskip}{4mm}
  A schematic drawing of the balance jig for lifting the detector.
It shows a top view (top), a front side view (bottom left), and the other side view (bottom right).
The dimensions are given in mm.
}
 \label{Fig:CraneJig}
\end{figure}

The 70 tonne detector with Gd-LS and LS needs to be transported using a crane of 130 tonne capacity
during the MLF maintenance period, as shown in Fig.~\ref{Fig:IL}. A balance jig is used for the crane work as shown in Fig.~\ref{Fig:CraneJig}.
It can withstand a load of 80 tonnes, sufficient for the filled detector.

\begin{figure}[h]
 \centering
 \includegraphics[width=0.8 \textwidth]{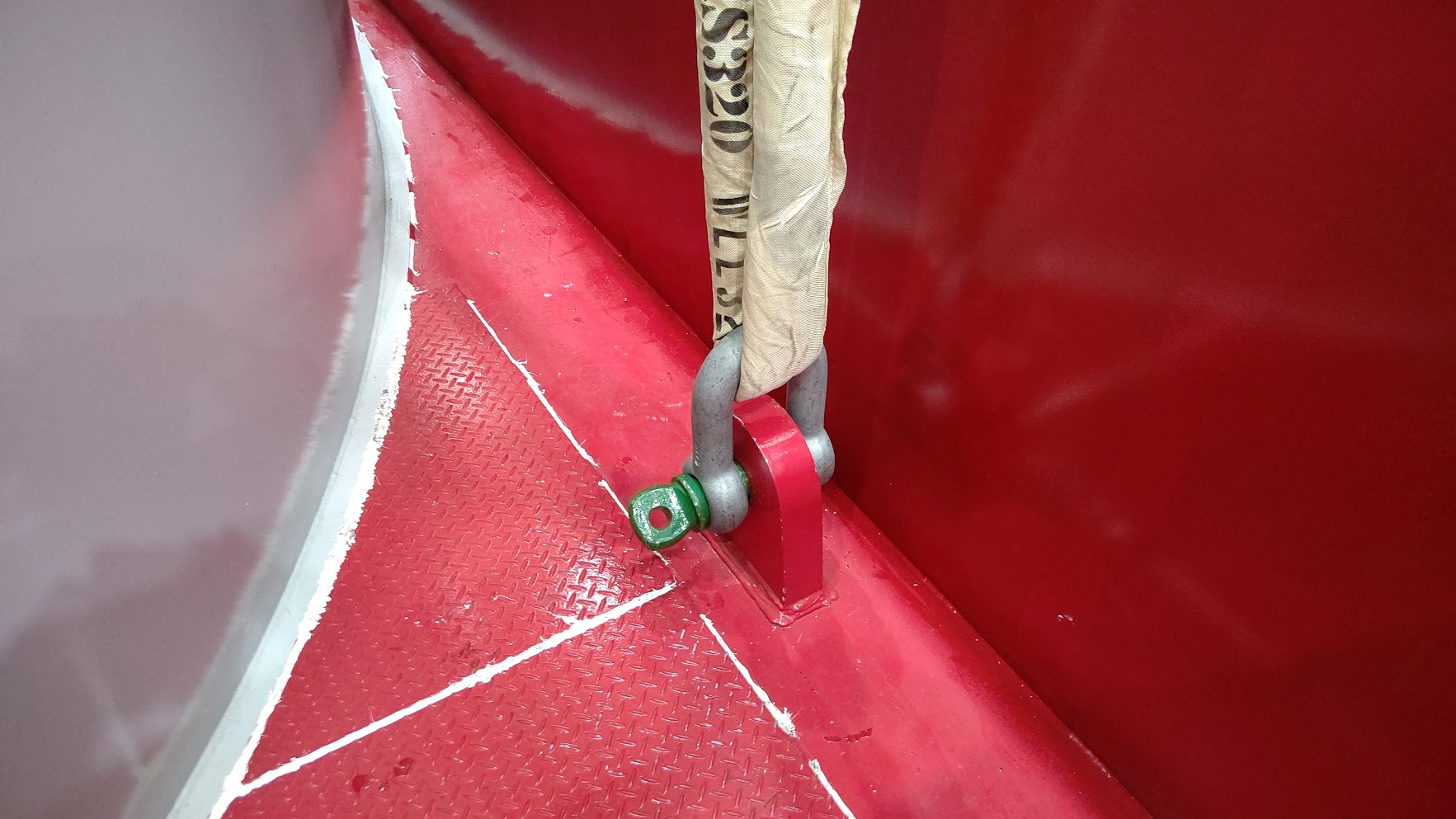}
\caption{\setlength{\baselineskip}{4mm}
  A shackle tied to the detector. Four shackles are used to lift the detector.
}
 \label{Fig:Shackle}
\end{figure}

Four slings are used to bind the jig and the detector together and connect them to the crane, as shown in Fig.~\ref{Fig:IL}.
The sling is tied with a shackle that is linked to a metal ring welded to the detector, as shown in Fig.~\ref{Fig:Shackle}.

\subsection{Acrylic Vessel}
\indent

\begin{figure}[htbp]
\centering
\includegraphics[width=0.7\textwidth,angle=0]{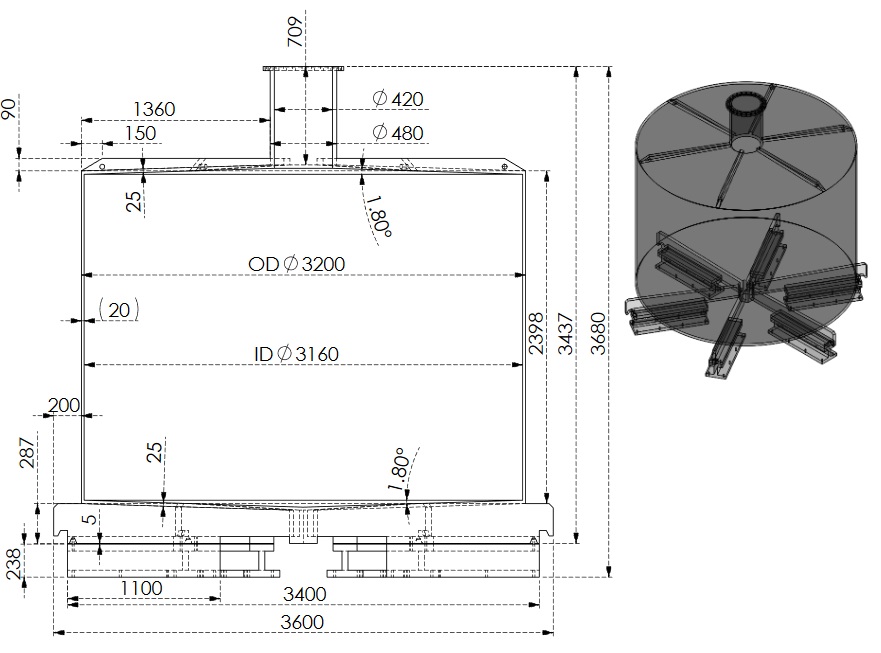}
\includegraphics[width=0.35\textwidth,angle=0]{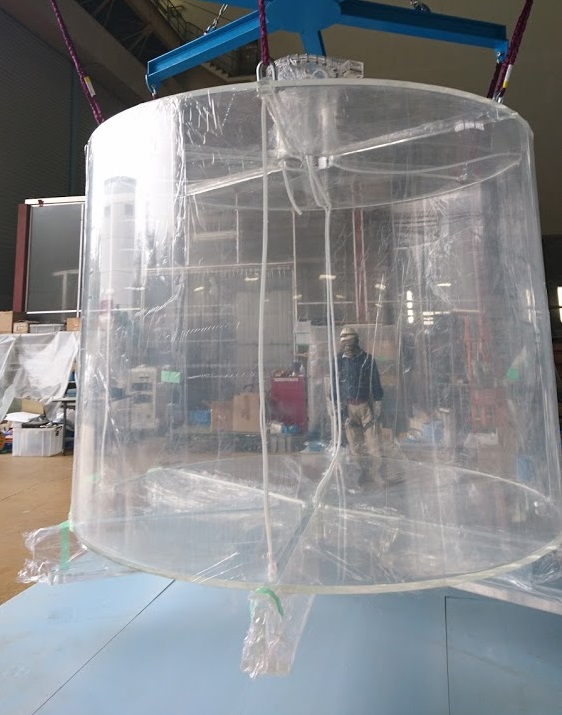}
\caption{\setlength{\baselineskip}{4mm}
    Schematic drawing (top) and picture (bottom) of the acrylic vessel.
The dimensions are given in mm.
}
\label{Fig:AV}
\end{figure}

A schematic drawing, as well as a picture, of the acrylic vessel are shown in Fig.~\ref{Fig:AV}.
A tapered design for its top and bottom plates is adopted to avoid air bubbles in the LS and
provide mechanical strength.

\begin{figure}[htbp]
\centering
 \includegraphics[width=0.60 \textwidth]{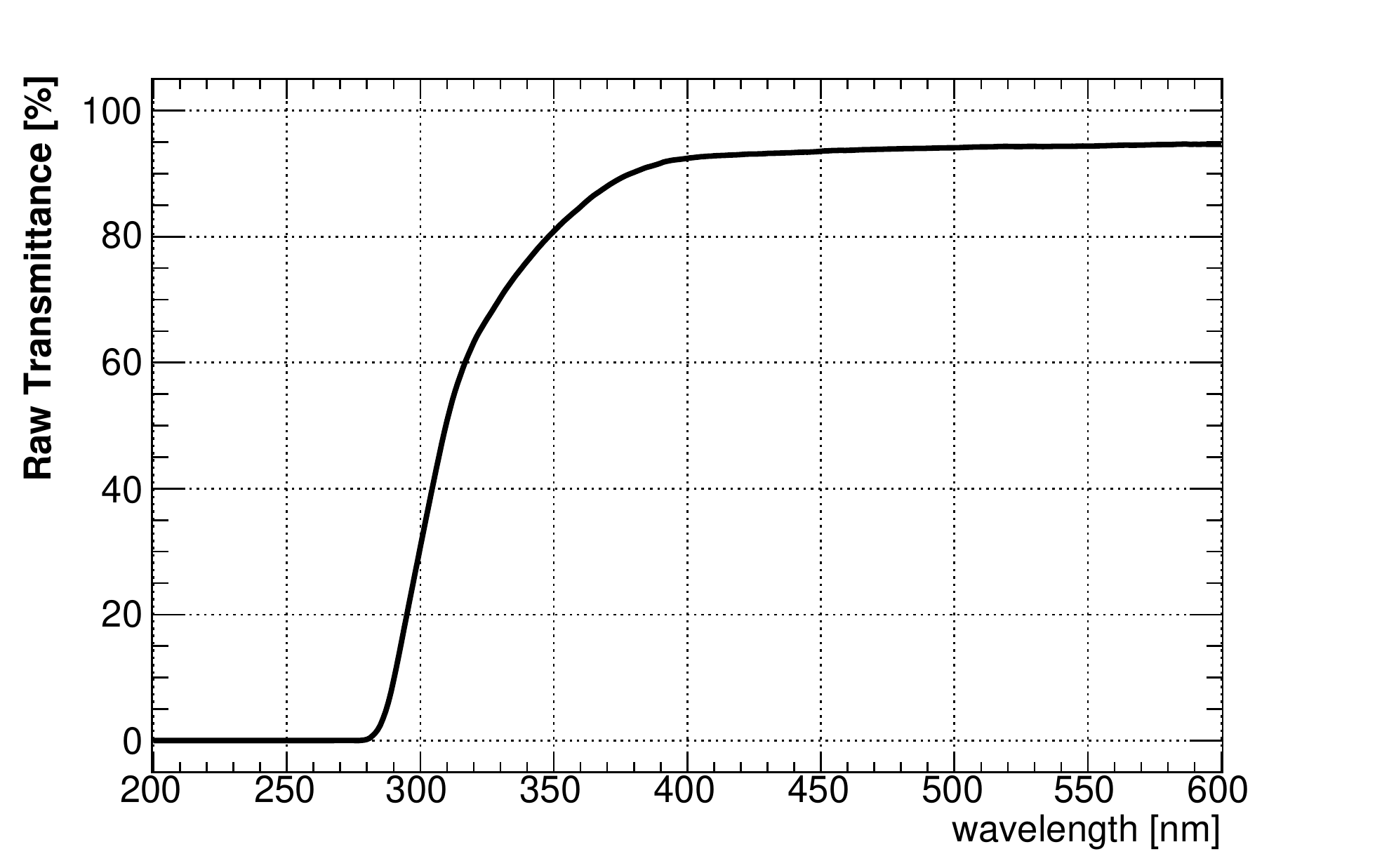}
\caption{\setlength{\baselineskip}{4mm}
  Measured transmittance of acrylic plates with 25\,mm in thickness.
The transmittance is better than 90\% at wavelengths longer than  400 nm.
}
\label{Fig:TPA}
\end{figure}

The thicknesses of the acrylic plates are 20, 25, and 30\,mm
for the barrel, top/bottom, and chimney parts, respectively.
Figure~\ref{Fig:TPA} shows measured transmittances of the 25\,mm thick
acrylic plates as a function of wavelength.
The UV transparent acrylic vessel was manufactured by the Nakano company in Taiwan in 2018 and
delivered to J-PARC in 2019. A leak test using nitrogen gas was successfully completed before delivery.
A more detailed description of the acrylic vessel can be found in Ref.~\cite{CITE:AV}.

\subsection{Veto}
\indent

The veto layer, filled with LS, surrounds the central target and gamma-catcher regions.
The thicknesses of the veto layers are 25\,cm at top and bottom, and 45\,cm at barrel.
The veto region is optically separated from the inner gamma-catcher by black acrylic boards.

\begin{figure}[htbp]
\centering

\includegraphics[width=0.45\textwidth,angle=0]{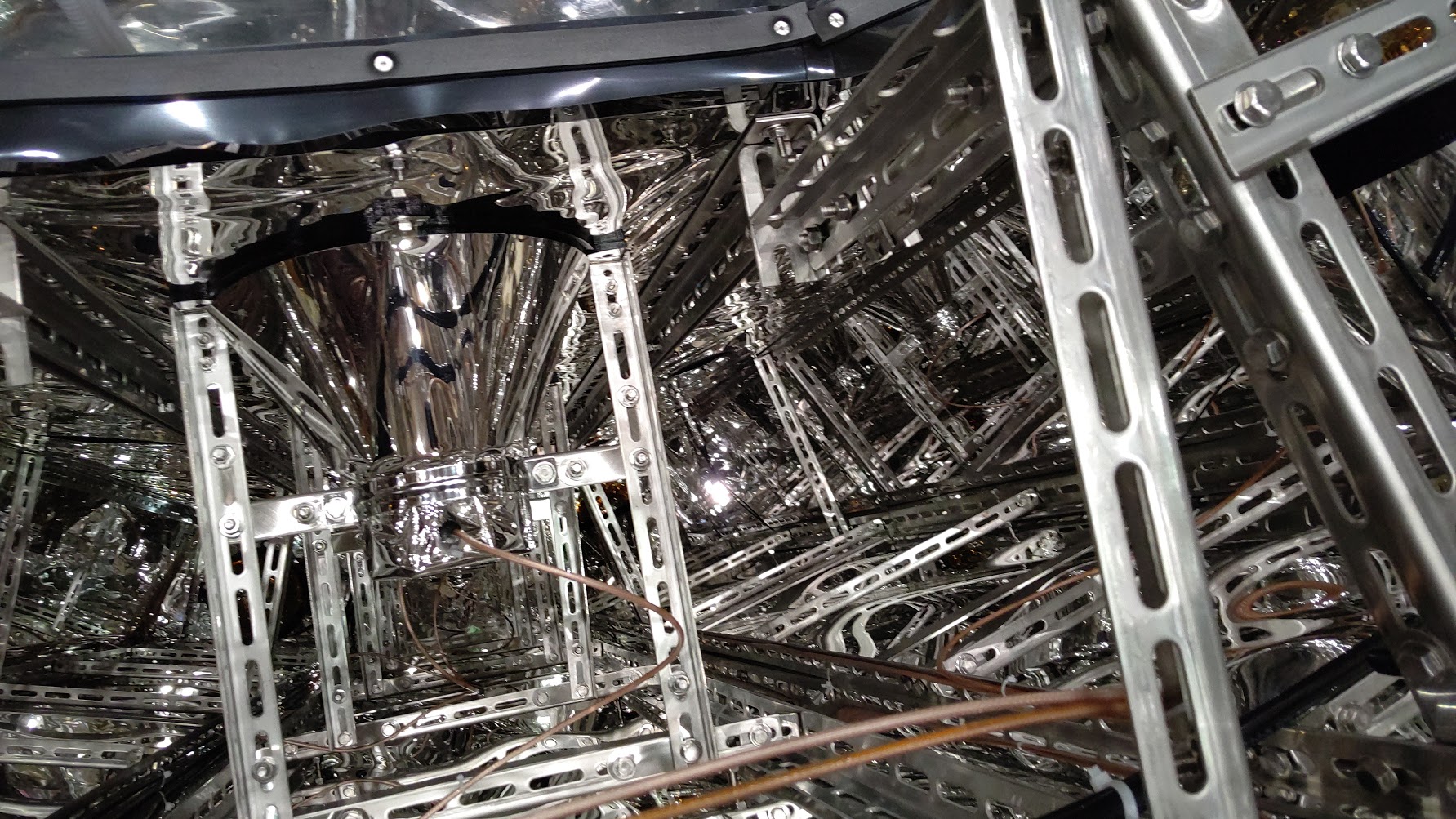}
\includegraphics[width=0.45\textwidth,angle=0]{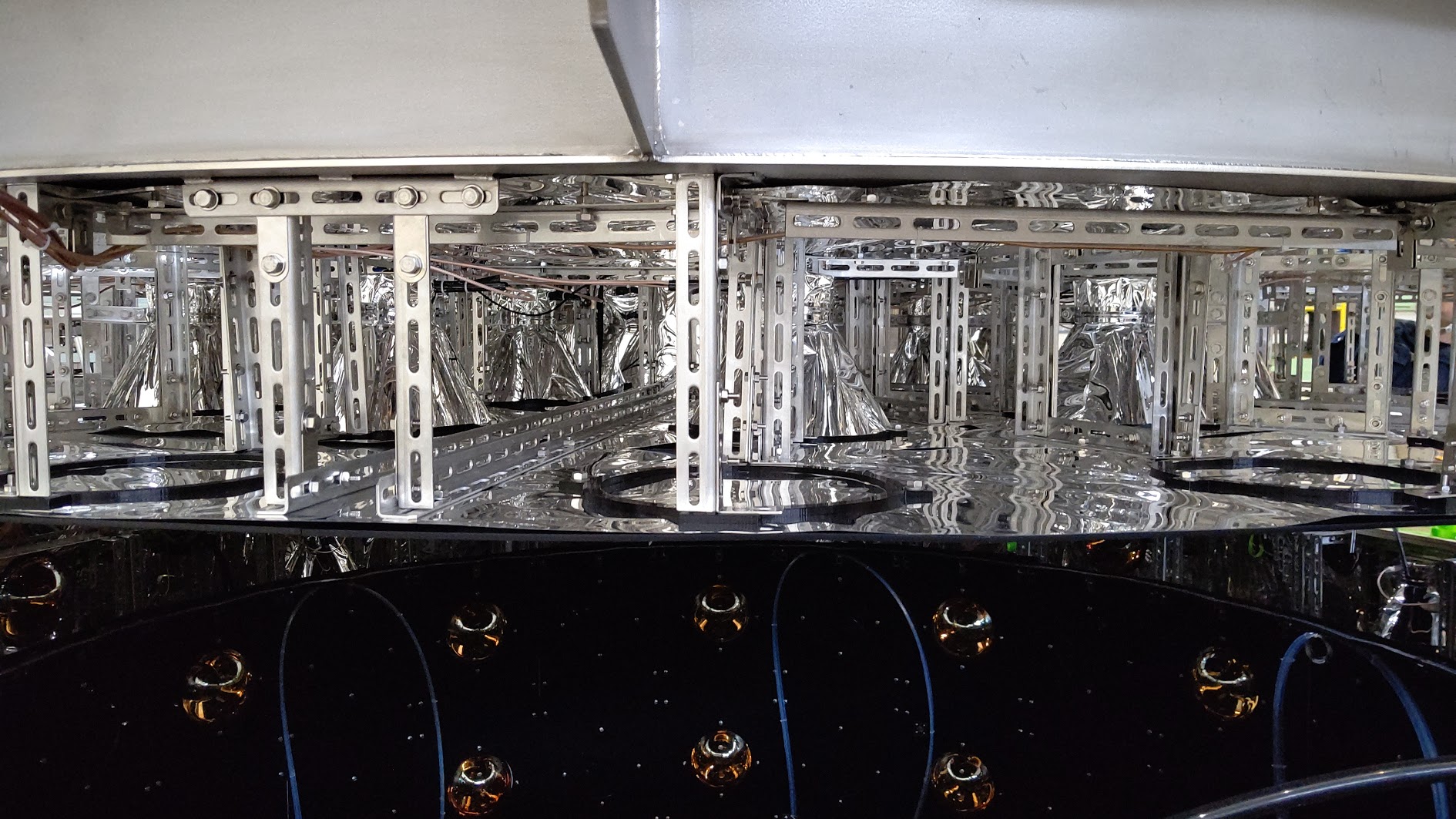}
\caption{\setlength{\baselineskip}{4mm}
  Pictures of the veto region, barrel (left) and top-lid (right).
}
\label{Fig:Veto}
\end{figure}

The veto layer is useful for identifying charged particles coming from outside the detector
and providing additional shielding from external backgrounds.
The surfaces of the veto layer are covered
with reflective sheets of REIKO LUIREMIRROR~\cite{CITE:REIKO}
to maximize the number of photons received by PMTs.
The measured reflectance of the sheet is greater than 94$\%$ for wavelengths longer than 440\,nm.
Figure~\ref{Fig:Veto} shows the veto region, with the reflective sheets clearly visible.

\subsection{Liquid Level Stabilization System}
\indent

The thermal expansion coefficient of Linear Alkyl Benzene (LAB) is
$\Delta\,V/V = 9 \times10^{-4}$ / $^{\circ}$C~\cite{CITE:DB}.
Since the maximum temperature variation of the experimental
area is expected to be $\pm$5 degrees, the liquid volume changes as much
as $\pm$0.45$\%$. The volume change leads to a change in the Gd-LS level in the acrylic chimney
by approximately 14\,cm/$^{\circ}$C. A liquid level stabilization system is therefore necessary.
Note that a change of the Gd unloaded LS level is minimized by the buffer space
along the top edge of the stainless steel tank as shown in Fig.~\ref{Fig:SStankDrawings}.

\begin{figure}[h!]
 \centering
 \includegraphics[width=0.55 \textwidth]{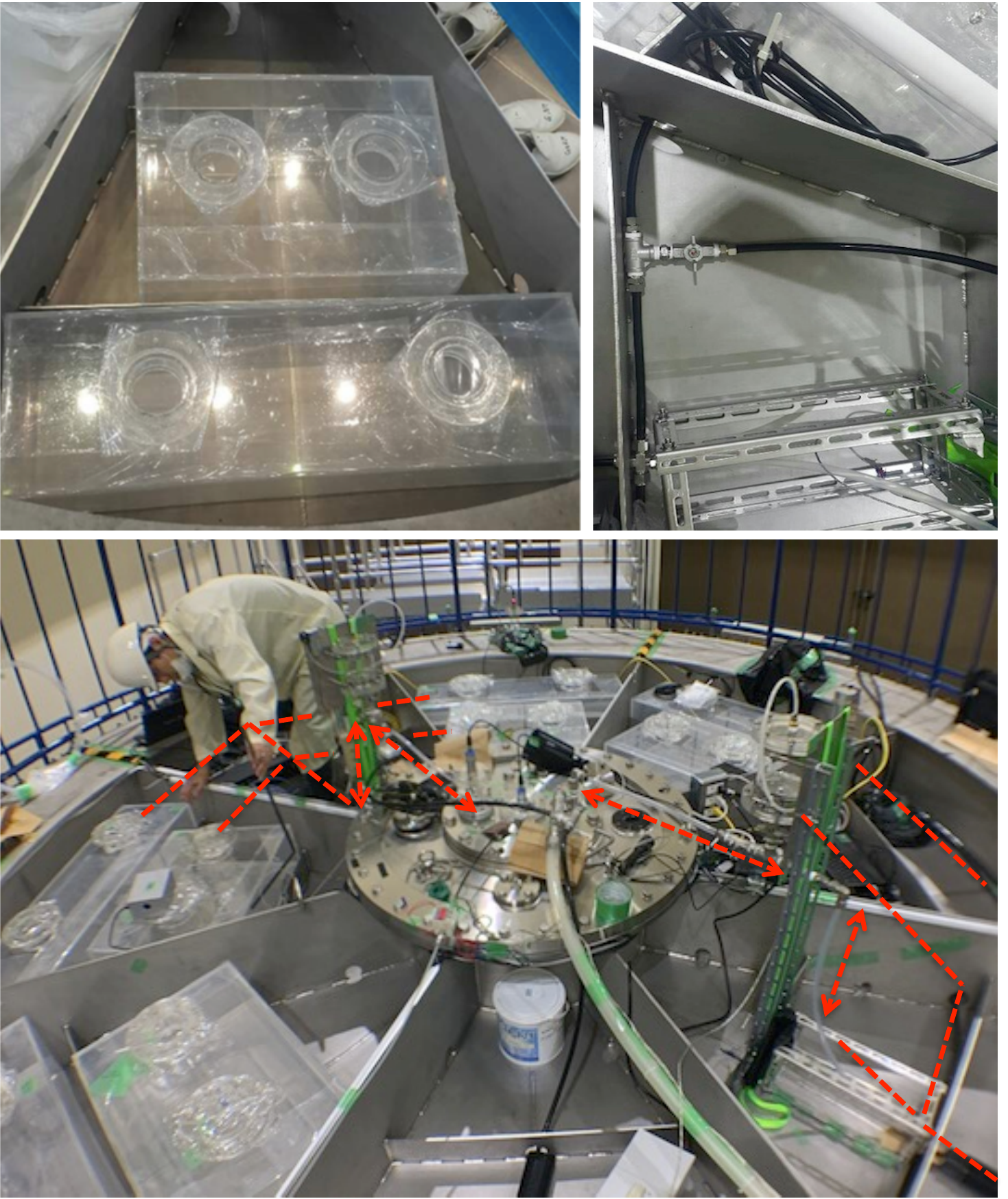}
\caption{\setlength{\baselineskip}{4mm}
Liquid level stabilization system consisting of eight acrylic containers and an inverse siphon. A pair of containers are deployed on each section of the detector lid (top left). The red dashed lines indicate pipe connections between the acrylic containers and to the siphon system.
}
 \label{Fig:LSV}
\end{figure}

A total of eight acrylic containers, four identical pairs, are mounted inside the detector lid as shown in Fig.~\ref{Fig:LSV}
to ensure a minimum change of the Gd-LS level and keep its level equal to the LS.
Each pair of two containers has a volume of 480 $\times$ 660 $\times$
240 mm$^3$ and 1100 $\times$ 330 $\times$ 240 mm$^3$.

\begin{figure}[h]
 \centering
 \includegraphics[width=0.6 \textwidth]{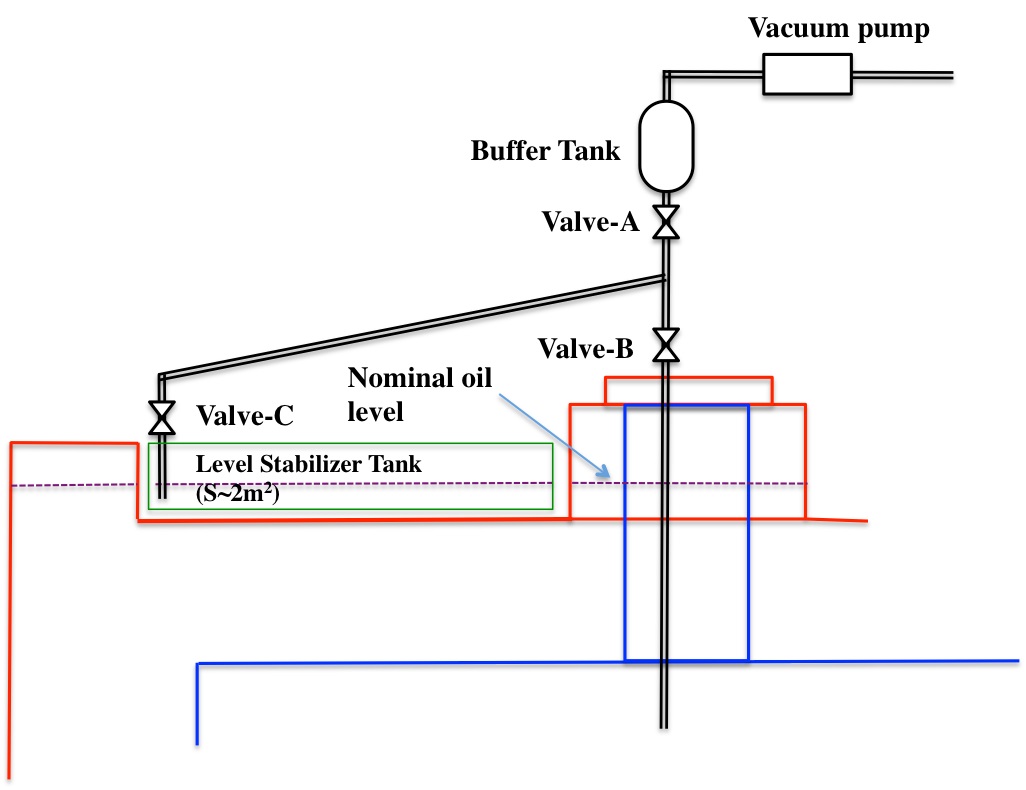}
\caption{\setlength{\baselineskip}{0mm}
Conceptual design of the liquid level stabilization system.
The nominal liquid level is above the top lid indicated by the dashed line. The vacant spaces in the upper corner of the stainless steel tank, stabilization acrylic containers, and chimney are filled with the nitrogen gas supplied from a generator at a flow rate of 3\,L/min.
}
 \label{Fig:siphon}
\end{figure}

Two sets of four containers are connected together by pipes and also to
the acrylic vessel inside the detector via an inverse siphon system, as shown in Fig.~\ref{Fig:siphon}.

The effective surface area of the Gd-LS is expanded from 0.14\,m$^2$ to approximately 2.5\,m$^2$ by this
system. As a result, the variation of the Gd-LS liquid level is only 0.7\,cm/$^{\circ}$C.
The acrylic stabilization containers and chimney are partially filled with Gd-LS. The stainless steel tank is also partially filled with LS, leaving a vacant space in the upper corner as shown by red circles in Fig.~\ref{Fig:SStankDrawings}. The remaining space is filled with nitrogen gas to prevent oxygen contamination from the outside. A positive pressure of 0.24\,kPa is maintained by the supplied nitrogen gas at a flow rate of 3.4 L/min.

\subsection{PMT Cable Flange}
\indent

\begin{figure}[h!]
 \centering
 \includegraphics[width=0.70 \textwidth]{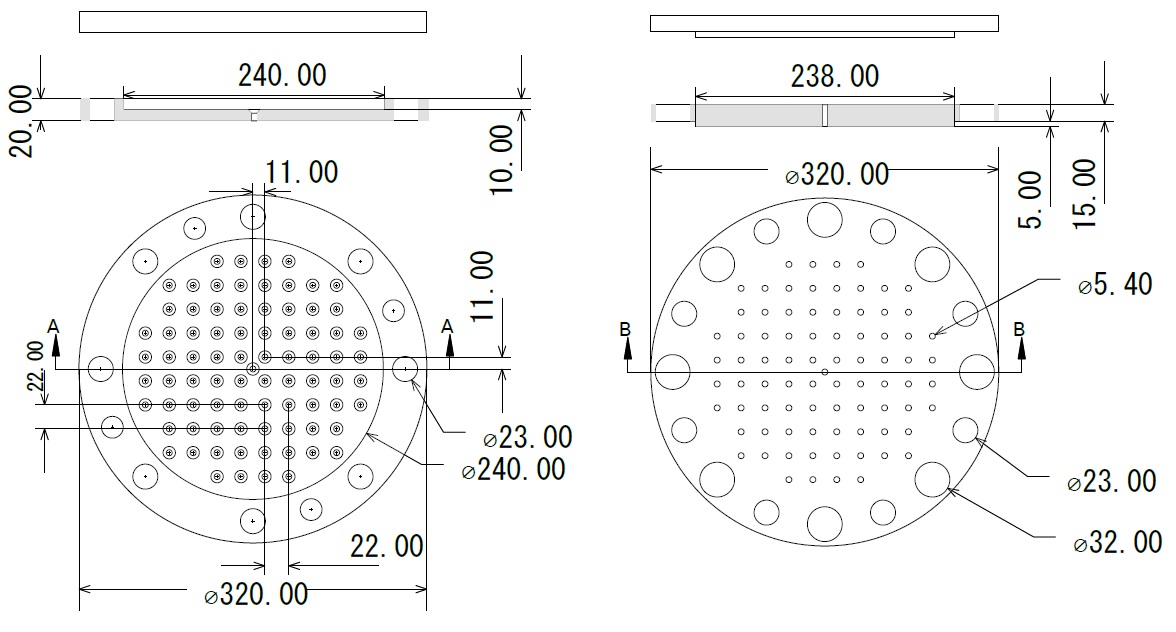}
\caption{\setlength{\baselineskip}{0mm}
  A schematic drawing of the PMT cable flange. A flange consists of two layers, a lower (left) layer and an upper (right) layer.
There are a total of 81 cable holes in the flange.
The dimensions are given in mm.
}
 \label{Fig:flange2}
\end{figure}

The PMT cables come out of the detector through four flanges.
The PMT cable flanges were especially designed to ensure air-tightness.
Figure~\ref{Fig:flange2} shows a schematic drawing of one of the flanges.

\begin{figure}[ht]
  \centering
  \includegraphics[width=0.3\textwidth,angle=0]{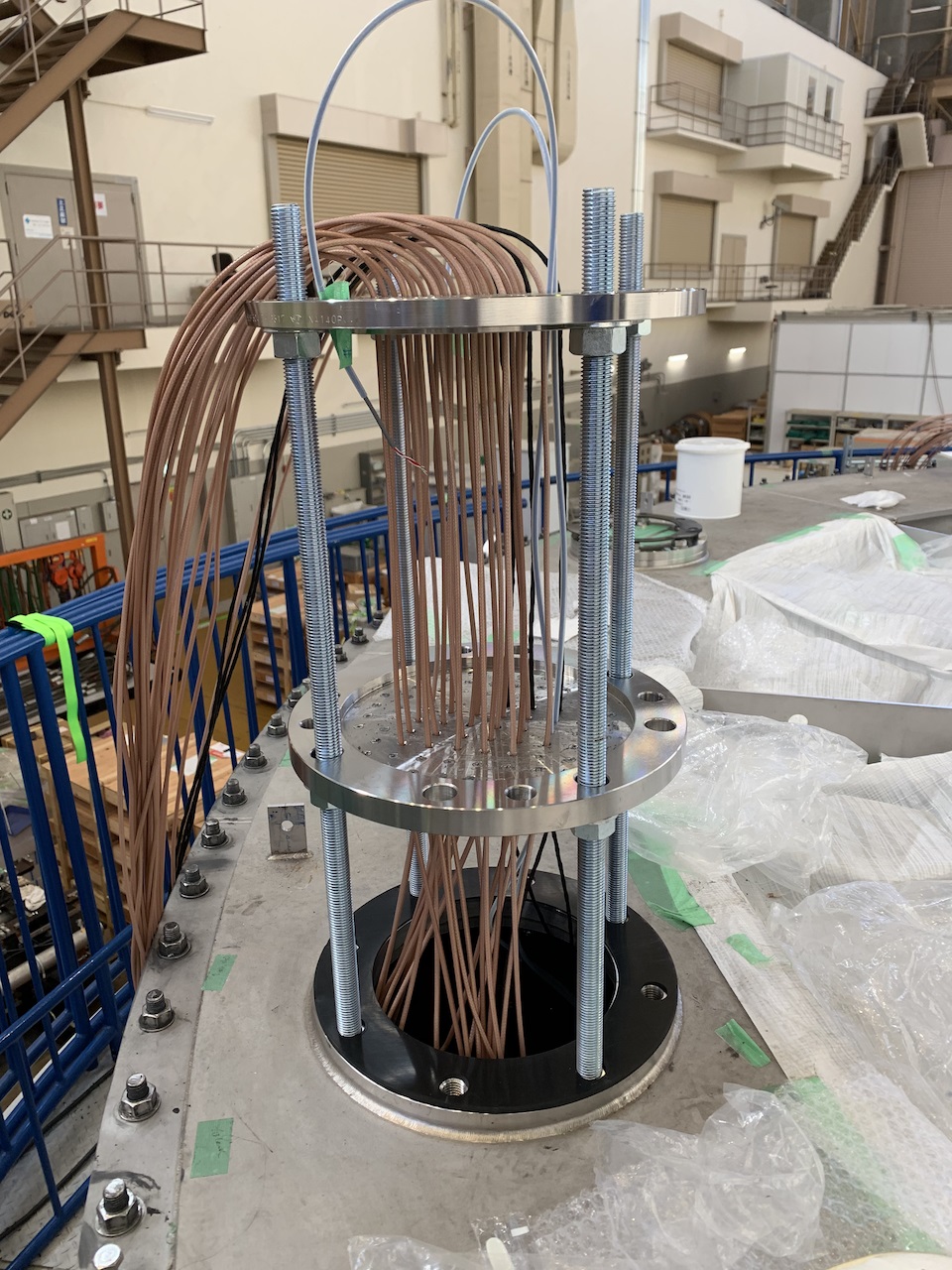}
  \includegraphics[width=0.3\textwidth,angle=0]{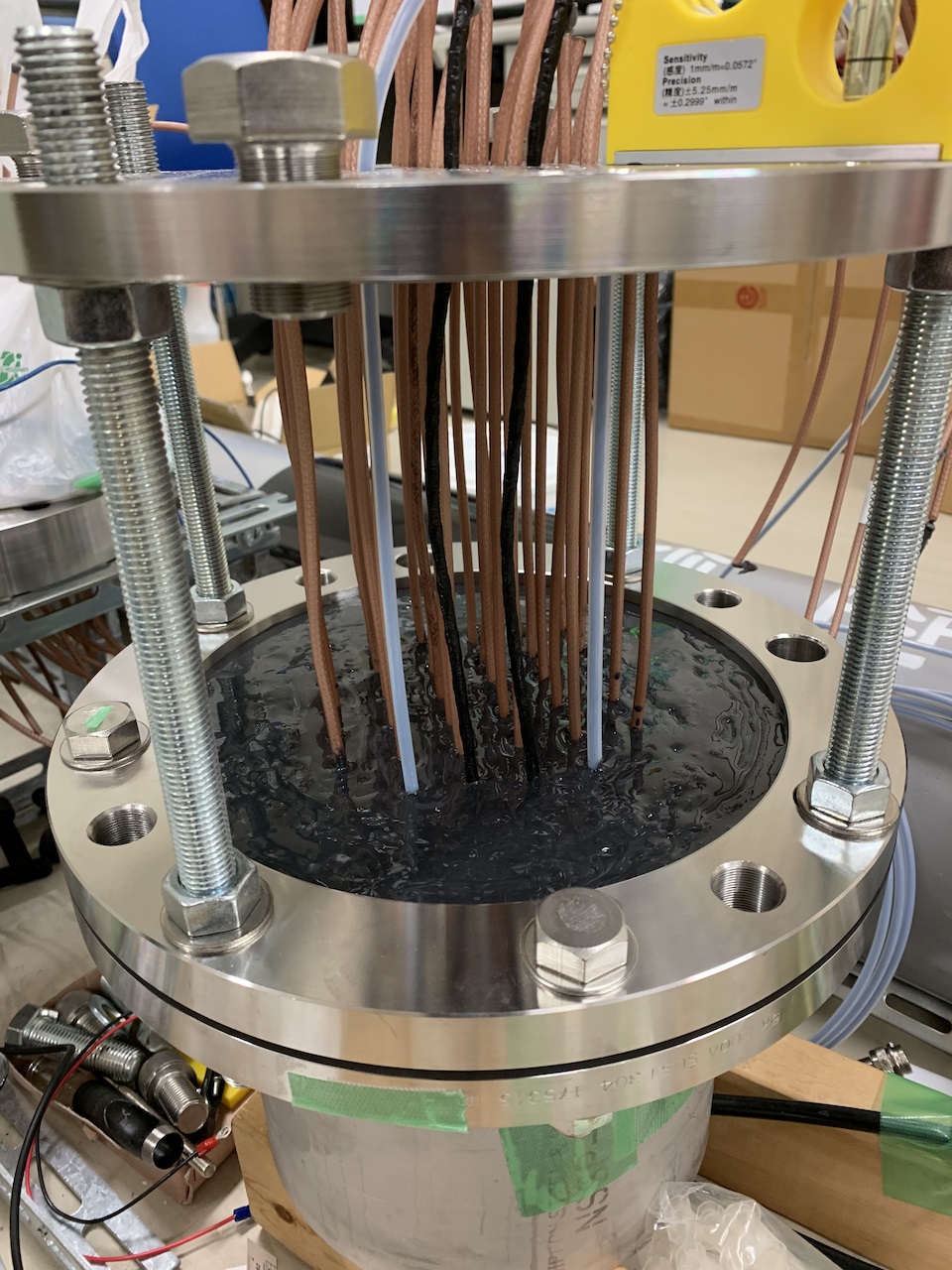}
  \includegraphics[width=0.3\textwidth,angle=0]{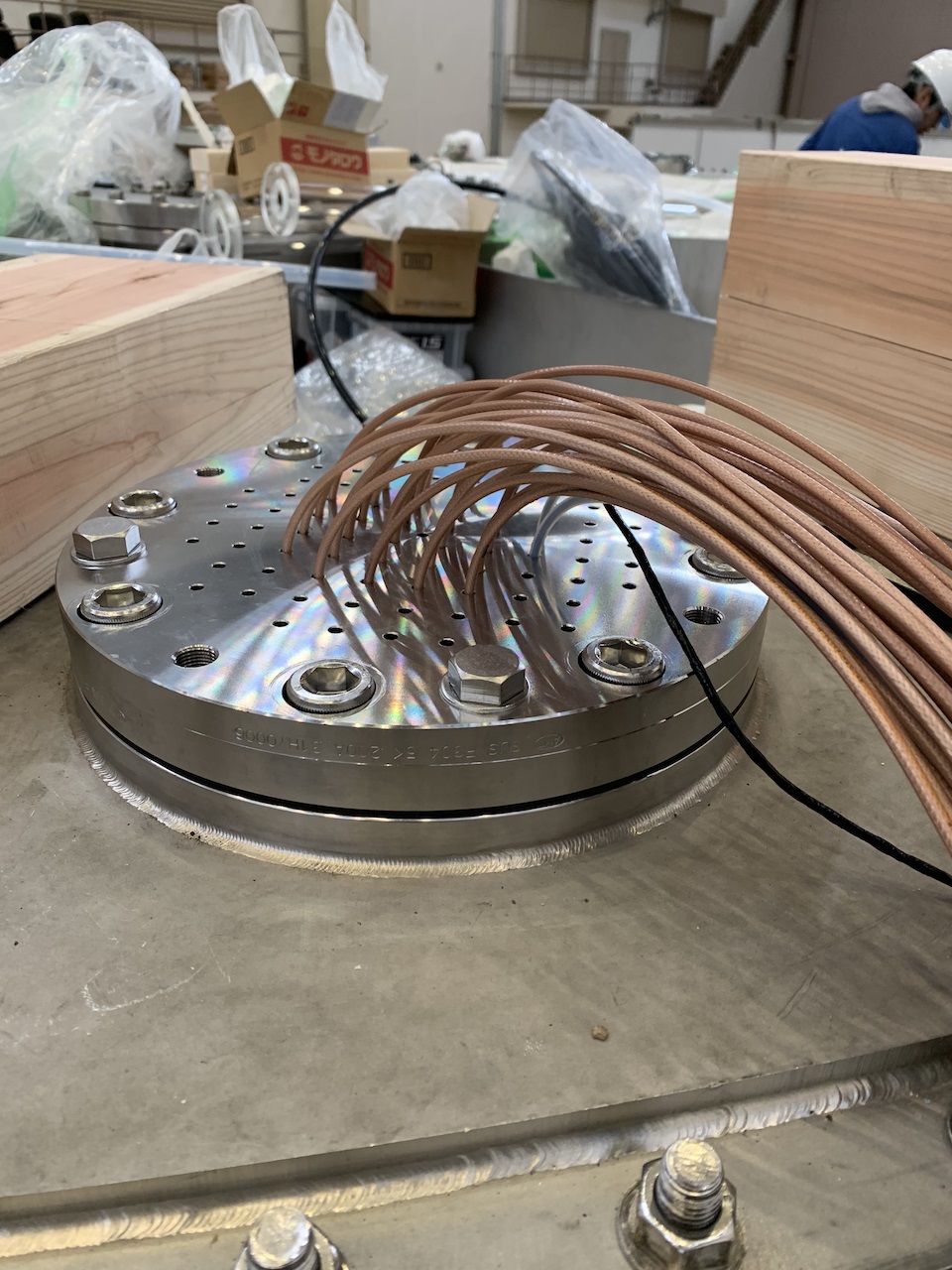}
  \caption{\setlength{\baselineskip}{4mm}
    PMT cable flanges. Left: A flange consists of double layers.
  Center: The lower layer is filled with the epoxy resin.
  Right: The top layer is used to close the flange.
  }
  \label{Fig:flange}
\end{figure}

Figure~\ref{Fig:flange} shows a flange with its PMT cables fed through.
The lower layer is filled with an epoxy resin after the PMT cables go through the flange holes
and the top layer is used to close the flange.

\section{Photomultiplier Tubes}

JSNS$^2$ employs 120 R7081~\cite{R7081, icecube} Hamamatsu 10 inch oil-proof PMTs, of which most were used previously by the RENO and Double Chooz experiments~\cite{RENO_pmt, DC_pmt}.
The photocathode material is a bialkali metal with a spectral response range between 300\,nm to 650\,nm and a maximum response at 420\,nm. The typical quantum efficiency at 390\,nm is 25\%. The characteristics of PMT R7081 are given in Table~\ref{tab:PMT}.

\begin{table}[h!]
\begin{center}
\begin{tabular}{c | c | c | c | c}
\hline
 Gain                            & Applied HV       & Quantum eff.           & Dark rate    & Peak to \\
                                   &  (V)                & at 390\,nm (\%)    & (kHz)        & valley ratio \\ \hline
1.0 $\times$ 10$^{7}$   & 1500               & 25                         & 7 $-$ 15   & 1.5 $-$ 2.8 \\ \hline
\end{tabular}
\end{center}
\caption{\setlength{\baselineskip}{4mm}Basic characteristics of the PMT R7081.}
\label{tab:PMT}
\end{table}

The high voltage (HV) is supplied to each PMT using CAEN A1535 modules~\cite{CAEN-HVmod} with the photocathode potential set to ground. A single coaxial cable is used for both high voltage supply and PMT signal readout. Double Chooz splitter boards, for decoupling the HV and PMT signal, are reused, and are described in more detail in Section~5.

A linear dynamic range of 1$-$500 photoelectrons (p.e.) is required to maintain full efficiency up to $\sim$60\,MeV,  the maximum prompt energy of an IBD event. The gain linearity is measured using laser and scintillation light. Typically, a PMT saturates at roughly 600 p.e. at $5 \times 10^6$ gain.

\begin{figure}[h!]
\centering

\includegraphics[width=60mm, height =55 mm]{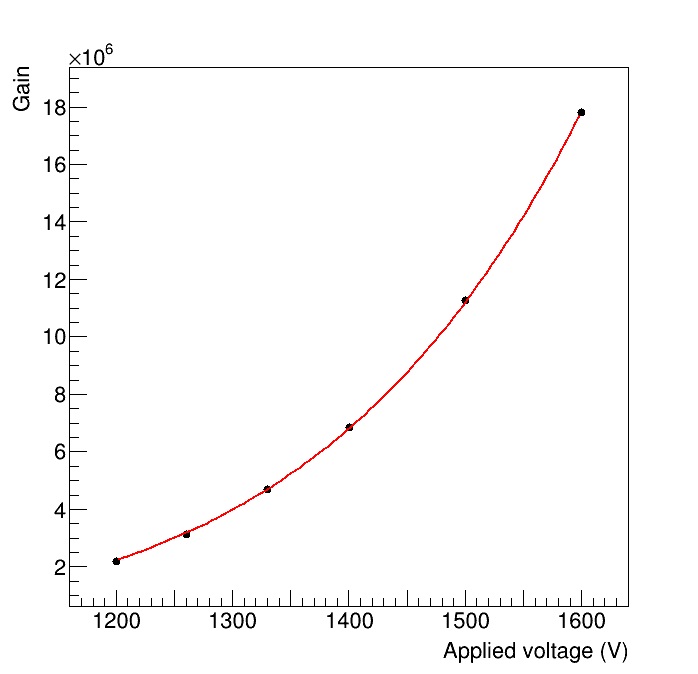}
\includegraphics[width=60mm, height =55 mm]{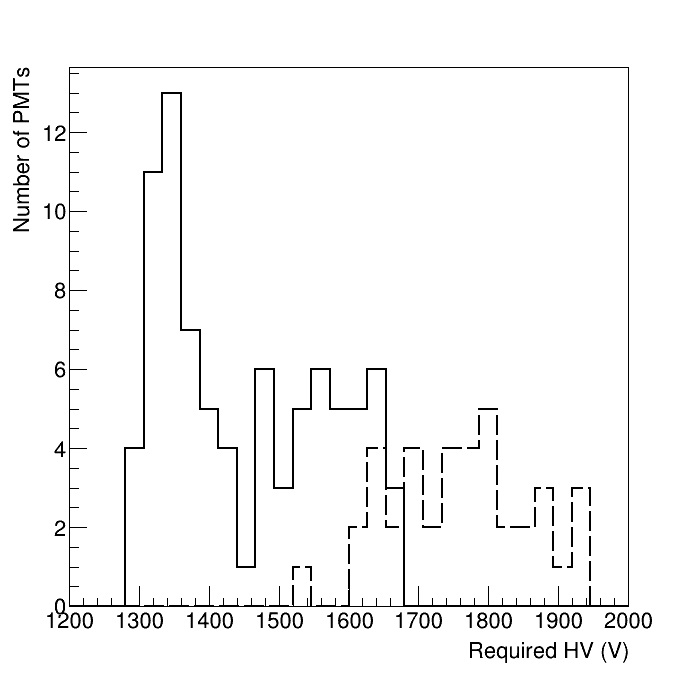}

\caption{Gain curve of a typical PMT (left) and HV required to achieve a gain of 10$^{7}$ (right). The solid histogram is for PMTs from RENO and Double Chooz and the dashed histogram for purchased new PMTs.  }
\label{gain}
\end{figure}

For the JSNS$^2$ experiment 40 PMTs were newly purchased, while the remaining PMTs were obtained from the RENO and Double Chooz experiments. All PMTs were carefully calibrated before being deployed in the detector.
For the PMT calibration, we used a 440\,nm nanosecond-pulse laser and a 14-bit ADC board to measure the gain, peak to valley ratio of the single photoelectron distribution, and the dark hit rate. A total of eight PMTs were simultaneously calibrated by using a light splitter and a dark box with 8 PMT mounting ports. The calibration apparatus was developed by Double Chooz~\cite{DC_pmt}. Figure~\ref{gain} shows the gain curve of a typical PMT, used to find the required HV for a gain of 10$^{7}$. The peak-to-valley ratios of the evaluated PMTs were found to be larger than 2 at that gain. The dark rate at a threshold of 0.3 p.e.\ was less than 3 kHz. The results obtained from this calibration are consistent with those  provided by Hamamatsu Photonics~\cite{JSNS_pmt}. Following our testing, 123 PMTs were identified to fulfill all quality criteria required for integration into the JSNS$^2$ experiment. The detailed description of the PMT calibration is available in  Ref.~\cite{JSNS_pmt}. Once installed in the detector, the timing performance of the PMTs is further calibrated using in-situ measurements provided by the LED system described in Section 9.

\begin{figure}[h]
\begin{center}
\includegraphics[scale=0.3]{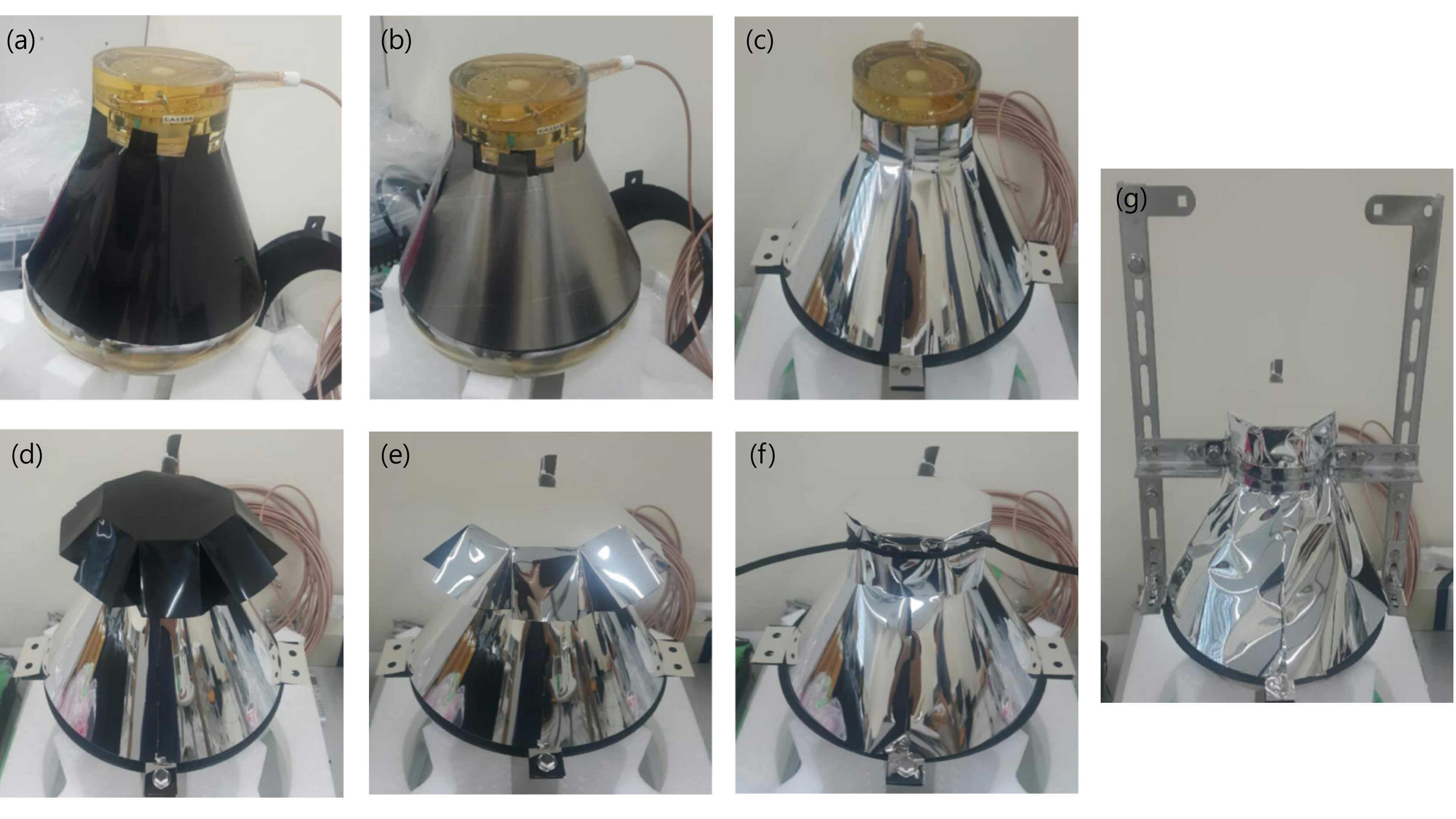}
\end{center}
\caption{\setlength{\baselineskip}{4mm}PMT assembly process consisting of (a) black PET, (b) FINEMET, (c) reflector, (d) black PET for PMT base and cable, (e) reflector, (f) fixing the reflector, and (g) L-type angle support frame.}
\label{PMT_assembly}
\end{figure}

A total of 120 PMTs are installed in the detector. Each PMT is assembled with an optical separation cover of black PET film, a magnetic shield of FINEMET~\cite{finemet1} and a cone-shaped reflection cover.
The black PET film~\cite{BlackPET} is used to reduce the background light coming from light flashes originating from inside of the PMT base circuit.
The FINEMET film improves the collection efficiency of photoelectrons by shielding the PMT from the geo-magnetic field~\cite{finemet2}.
The veto wall and the backside of inner PMTs are covered by reflective sheets of REIKO LUIREMIRROR with a reflectance of 94\% at 440\,nm wavelength~\cite{CITE:REIKO}. Figure~\ref{PMT_assembly} shows the PMT assembly process before mounting it on the detector.

\begin{figure}[h!]
\centering
\includegraphics[width=60mm, height =60 mm]{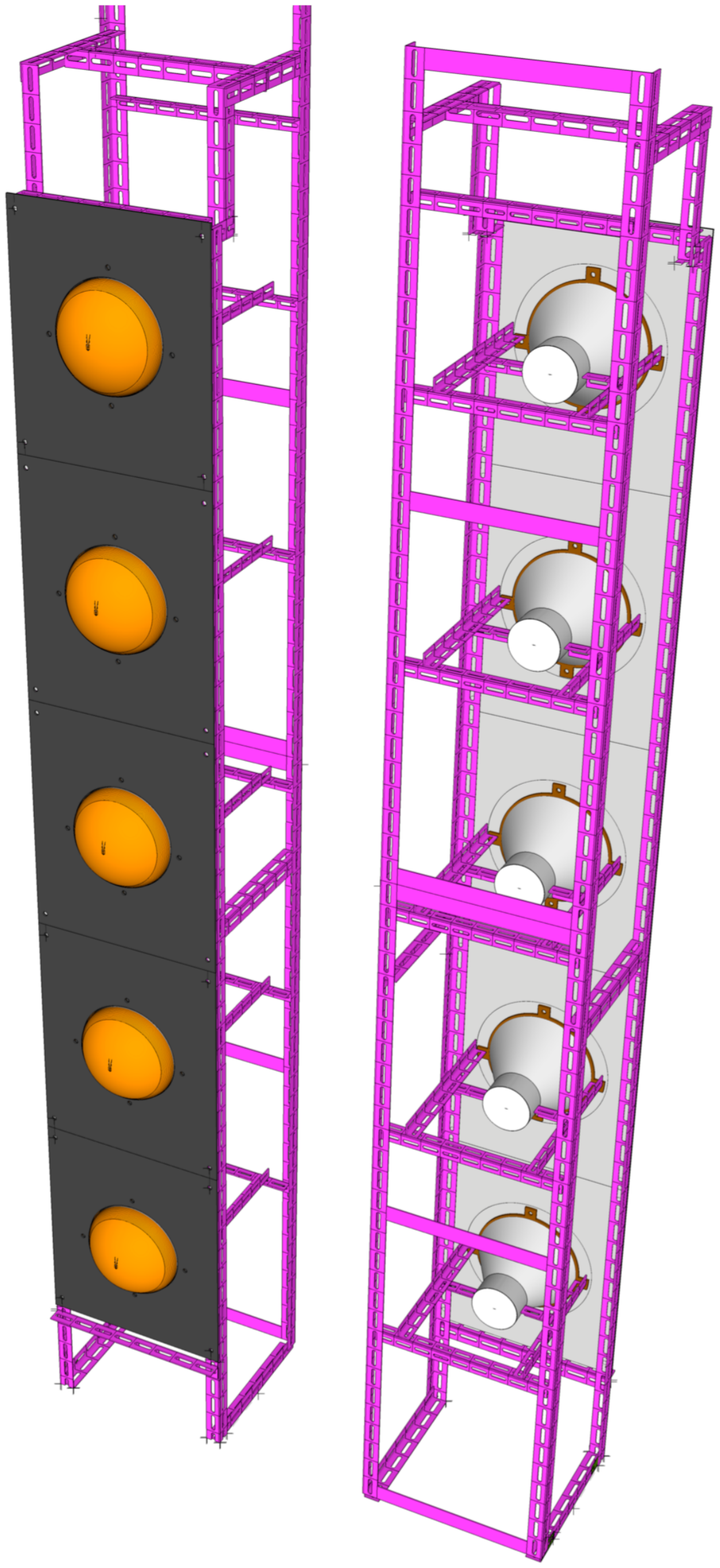}

\caption{Front (left) and backside (right) views of a barrel PMT support structure module.}
\label{PMT_mounting}
\end{figure}

An assembled PMT is mounted to a SUS L-type angle support frame that is welded to the inner detector wall, as shown in Fig.~\ref{PMT_mounting}. A black acrylic board is attached to the PMT support structure for optical separation between the gamma-catcher and veto.
At the detector barrel side, five PMTs are mounted to a barrel layer of support frame unit. In total, 60 PMTs are mounted to twelve barrel support frame units.
At the detector top and bottom sides, 18 PMTs are mounted to two circular layers of the support frame unit.
For light detection in the veto region, 24 PMTs are installed in the upper and lower corners of the cylindrical detector.
Twelve PMTs are mounted on the detector top and directed downward, 
and the other twelve are mounted on the detector bottom circumference with six of them 
directed radially inward and the remaining six directed upward.

\section{Liquid Scintillator}

The JSNS$^2$ LS needs to differentiate the electron antineutrino signal from accidental and neutron backgrounds. It must produce sufficient light to allow for precise energy measurement, such that the spectral modulations expected from sterile neutrino oscillations can be observed. It must also have appropriate chemical properties to satisfy the Japanese fire law and maintain stable detection efficiency for several years.

To satisfy these requirements, the JSNS$^2$ experiment uses 31 tonnes of unloaded LS, produced by the RENO experiment's refurbished facility~\cite{JSNS2-LS}.
ISO (International Organization for Standardization) tanks, made of stainless steel (SUS316), are used for LS storage and transportation from the RENO site in Korea to the J-PARC MLF in Japan.
A detailed description of the LS production and its optical properties is given in Ref.~\cite{JSNS2-LS}.

The JSNS$^2$ LS is composed of LAB (${\rm C_n H_{2n+1}}$-${\rm C_6 H_5}$, $\mathrm{n} = 10\sim13$) as a base solvent, 3 g/L of PPO (2,5-diphenyloxazole) as a primary fluor, and 30 mg/L of bis-MSB (1,4-bis(2-methylstyryl) benzene) as a secondary wavelength shifter. LAB is chosen as an organic solvent because it is environmentally friendly,  nontoxic, and has a high flash point (152 \textdegree{}C).
It also has a long attenuation length, $>$10\,m at 430\,nm, and has a light yield of $\sim$8,000 photons per MeV~\cite{RENO-LS}. The Isu Chemical Company in South Korea produces high quality LAB and can deliver it in a clean storage container.
LAB suffers from a non-linear energy response due to quenching effects at low energies. The RENO experiment measured the Birk's constant for LS as 0.0117$\pm$0.0003\,cm/MeV using a Ge detector~\cite{RENO-LS}.

The JSNS$^2$ experiment uses 17 tonnes of Gd-LS, as a neutrino target, which was donated by the Daya Bay experiment~\cite{DB-LS}. The IBD reaction is utilized to detect antineutrino appearance from sterile neutrino oscillations. A neutron coming from the IBD reaction captures on either Gd or hydrogen (H). While the neutron capture on H produces a 2.2 MeV gamma-ray, which cannot be distinguished  from natural radioactive backgrounds, the $\sim$8 MeV gamma-rays from the neutron capture on Gd can be easily identified above the background. The mean neutron-capture time on 0.1\% Gd is $\sim$30~$\mu$s, which is more than 6 times shorter than that on H, and can result in significant reduction of accidental backgrounds.

In accordance with Japanese fire law, the MLF has stringent safety requirements for how the flammable LS and Gd-LS must be treated.
To comply with these safety requirements, JSNS$^2$ has taken precautions to ensure the scintillator cannot spill or leak from the detector.
The temperature of the Gd-LS is monitored by eight sensors installed inside the detector.
The liquid level in the detector is monitored by ultrasonic sensors.
Several acrylic containers are mounted on top of the detector and provide a buffer volume for the Gd-LS.
The buffer volume helps maintain a constant liquid level in the detector, and prevents overflow as the liquid volume changes due to thermal expansion.
The liquid levels in the containers are also monitored by ultrasonic sensors. The detector is surrounded by a spill tank; any liquid spill can be observed by ultrasonic sensors and web cameras. Section 7 contains a more detailed description of the LS and Gd-LS monitoring.

A liquid handling system has been developed to complete liquid filling or extraction within a few days.
It consists of two pumps, several flexible pipes, two frequency inverters, and two flow meters.
An FM3104-PD-XP/K flow meter~\cite{flowmeter} is used to measure the liquid flow rate into the detector. The frequency inverter is used to adjust the pump speed to modulate the flow rate.
The liquid filling and extraction takes place at the first floor of the MLF and are both carefully performed to maintain similar Gd-LS and LS levels by controlling liquid flow rates with the inverter.
Three ISO tanks with heat insulation are used for the liquid storage and transportation.

\begin{figure}[hbt]
\begin{center}
\includegraphics[width=0.8\textwidth]{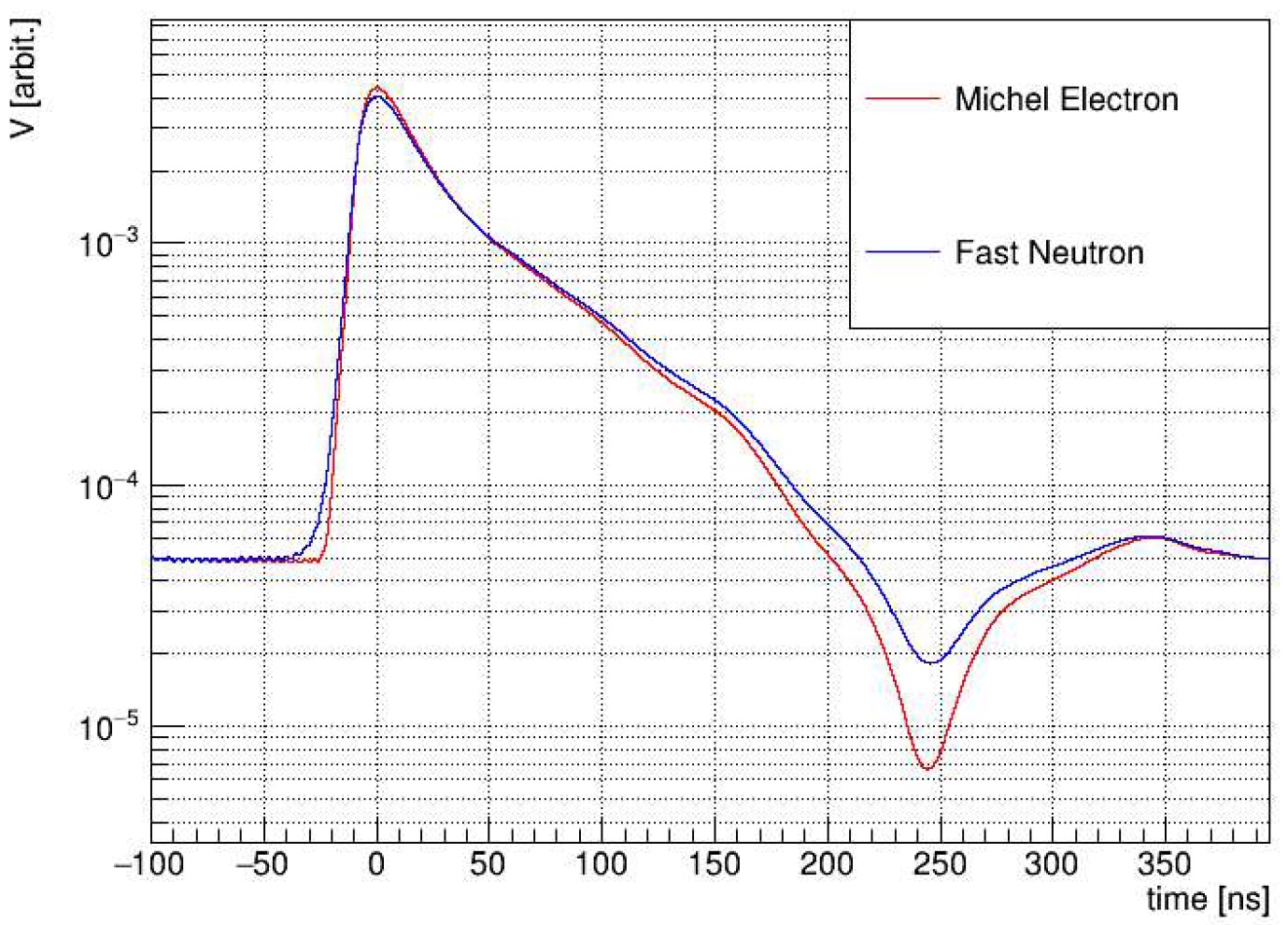}
\caption{Comparison of the average waveforms for fast neutron and electron events. The fast neutron’s average waveform is wider than that
of an electron.}
\label{PSD-waveform}
\end{center}
\end{figure}

One of the most serious backgrounds for the sterile neutrino search comes from fast neutrons produced by cosmic muons. An energetic neutron loses its energy through interactions with protons in the scintillator, and the recoil protons can produce sufficient scintillation light to fake an IBD prompt signal event. The neutron eventually thermalizes and captures on Gd or H. Thus, a fast neutron can mimic an IBD prompt and delayed signal. The JSNS$^2$ detector, not being underground, suffers significantly from the cosmic muon induced fast neutrons.
The charge waveform of a scintillation signal generated by a recoil proton is wider than that of a positron due to difference in the dE/dx between the two particles. By exploiting this difference in waveforms, the fast neutron background can be distinguished from an IBD prompt event using a pulse shape discrimination (PSD) algorithm. In order to improve the PSD performance, an additional solvent, DIN (di-isopropylnaphthalene), EJ-309 by Eljen~\cite{EJ-309}, was dissolved into the Gd-LS before the second run, in December 2020. The DIN concentration is 8\% of the Gd-LS, and an additional 2\% will be dissolved in July 2021. The different charge waveforms for neutron and electron interactions are shown in Fig.~\ref{PSD-waveform}. It is expected that 90\% of the fast neutron background can be removed using PSD.

\begin{figure}[hbt]
\begin{center}
\includegraphics[width=1.0\textwidth]{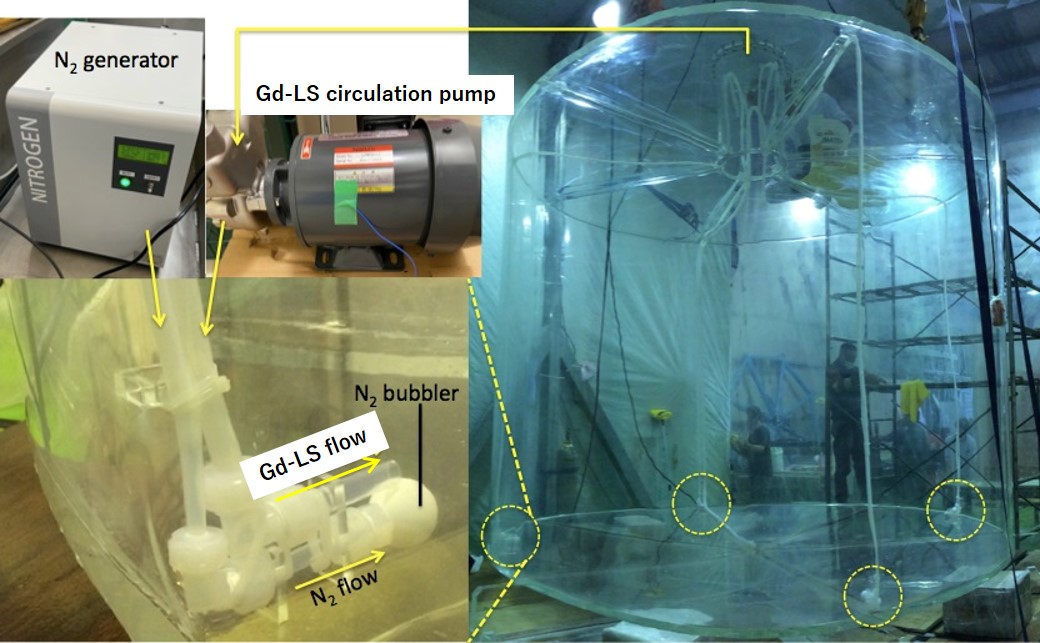}
\caption{Nitrogen purging system. Four Teflon bubblers are installed at the bottom corner of the inner acrylic vessel, indicated by the yellow circles (right). The nitrogen gas is provided by a generator (top left) through a Teflon tube and injected into Gd-LS via the bubbler holes (bottom left). The target liquid is circulated by a turbopump. The yellow solid arrows indicate the nitrogen gas flow and the Gd-LS circulation. }
\label{N2-purging}
\end{center}
\end{figure}

Nitrogen purging is used to remove oxygen and moisture from the Gd-LS. Oxygen in the Gd-LS causes degradation of its fluorescence and transparency. The nitrogen purging system consists of spherical Teflon bubblers (35\,mm diameter) with many 10\,$\mu$m size holes and a Cosmo CS MicroMini~1 nitrogen generator~\cite{N2-generator}. The bubblers are installed at the bottom corner of the inner target vessel, as shown in Fig.~\ref{N2-purging}. The generator outside the detector provides nitrogen gas at a rate of 0.8\,L/min, through a 3/8” Teflon tube attached to the vessel wall. During nitrogen purging, a NIKUNI 32NPD07Z nitrogen-regenerative turbopump~\cite{NIKUNI-pump} is used for circulating the Gd-LS at a rate of 28\,L/min. The liquid is extracted at the chimney and supplied back to the target volume, next to the bubbler, through a 1/2” Teflon tube. When the nitrogen purging system runs, the oxygen concentration in the chimney air increases because the dissolved oxygen is expelled from the Gd-LS. The oxygen in the chimney is removed by the nitrogen gas flow. Roughly five days of nitrogen purging and Gd-LS circulation removes most of the oxygen in the Gd-LS and increases the observed light yield.

\section{Electronics}

\begin{figure}[htbp]
\centering
\includegraphics[width=0.9\textwidth]{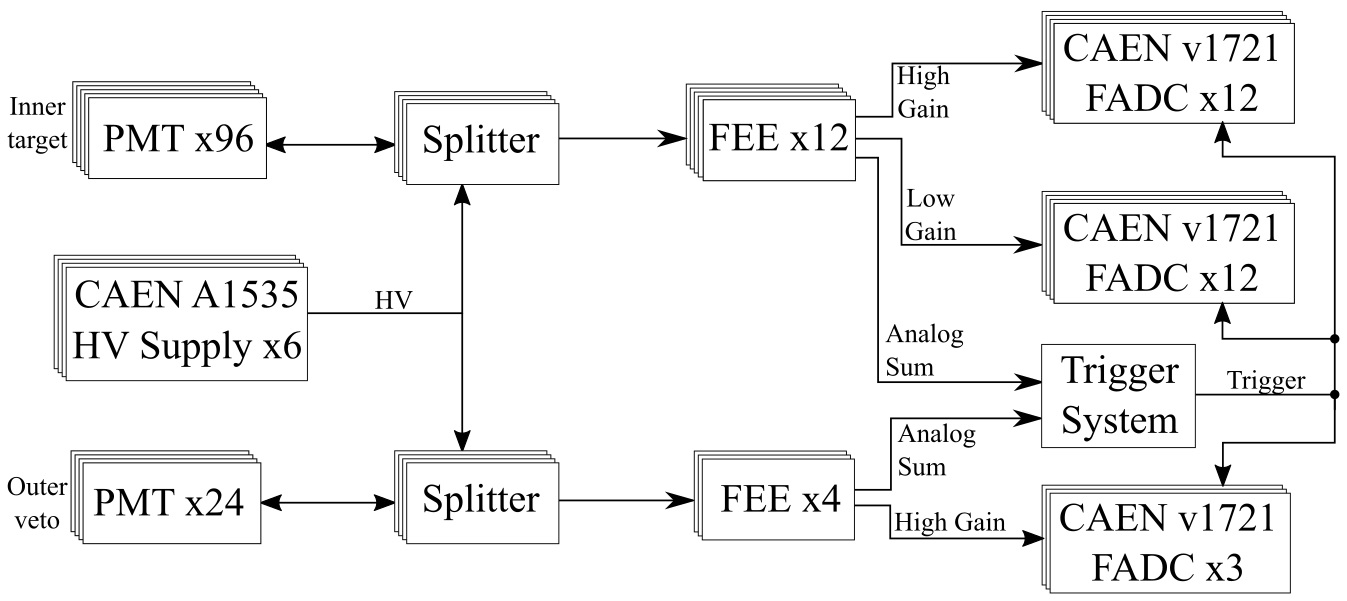}
\caption{Schematic overview of the JSNS$^2$ front end and DAQ electronics. The ``Analog Sum'' signal for the trigger system is described in Section 6.}
\label{front_end_diagram}
\end{figure}

A schematic overview of the DAQ system is shown in Fig.~\ref{front_end_diagram}.
The HV supply modules, HV splitter components, front-end analog processing
board and digitization modules were all reused from the Double Chooz experiment.

HV is supplied to each PMT individually from six CAEN A1535 modules, all housed within a purchased CAEN SY4527LC power supply crate~\cite{CAEN-HV}.
Each A1535 module supplies HV up to 3.5\,kV to 24 PMTs, each with a maximum current of 3\,mA.
The read-back voltage, supply current, and temperature of each of the modules are monitored as part of the JSNS$^2$ slow control and HV monitoring system.

The next component of the DAQ system is a HV splitter, responsible for separating the high speed PMT signal from the DC HV.
The splitter was developed by CIEMAT for the Double Chooz experiment.
The signal is separated from the HV using a single 6.8\,nF capacitor, rated up to 2500\,V.
An additional noise-filter circuit is included as well. The circuit diagram is available in Ref.~\cite{JSNS_pmt}.
The splitter circuit board is housed within a grounded metal box. Each splitter handles a single PMT.

The analog PMT signal from the splitter is further processed by Front-End Electronics (FEE).
The FEE modules were originally developed by Drexel University and Lawrence Livermore National Lab for the Double Chooz DAQ system.
Each FEE board processes eight PMT signals. It produces as output a high-gain and low-gain copy of each input signal, and also provides an analog sum of high gain outputs.

We introduced one significant modification to the FEE compared to the Double Chooz experiment.
The gain of the low-gain signal path is modified to reduce the output signal amplitude by a factor of $3.9$ compared to
the Double Chooz configuration.
The modification is made to increase the dynamic range of the DAQ system to achieve an increased sensitivity to high energy interactions.
All FEE measurements and values discussed here include that modification.

Twenty-seven CAEN V1721 FADC modules~\cite{CAEN-FADC} provide digitization and readout for PMT signals after the FEE's processing.
These modules digitize signals at 500\,M samples per second across a 1\,V dynamic range with 8-bit precision.
Each FADC module receives a trigger signal and a global synchronization clock from a special trigger FADC.
Each FADC module has an internal 125\,MHz counter that is used to reconstruct the time of an event relative to the other events
and other FADCs.
The FADCs are all readout by a single DAQ PC over an optical fiber connection.

The effective gain and dynamic range of the FEE's high and low gain signal path is of primary importance for JSNS$^2$.
The dynamic range limits our sensitivity to high energy interactions and precise knowledge of the gain is
necessary for reliable energy reconstruction.
Measurements have shown that the high gain channels saturate at an output voltage of
approximately $1700$\,mV with a corresponding input voltage of approximately $110$\,mV.
Similar measurements for the low gain channels showed saturation at an output voltage of $1500$\,mV
and a corresponding input voltage of approximately $3000$\,mV.
For the typical PMT gain in the JSNS$^2$, the electronics saturate at approximately 20 and 550 p.e.\ for the high gain and low gain, respectively.

\begin{figure}[htbp]
    \centering
    \includegraphics[width=0.48\textwidth]{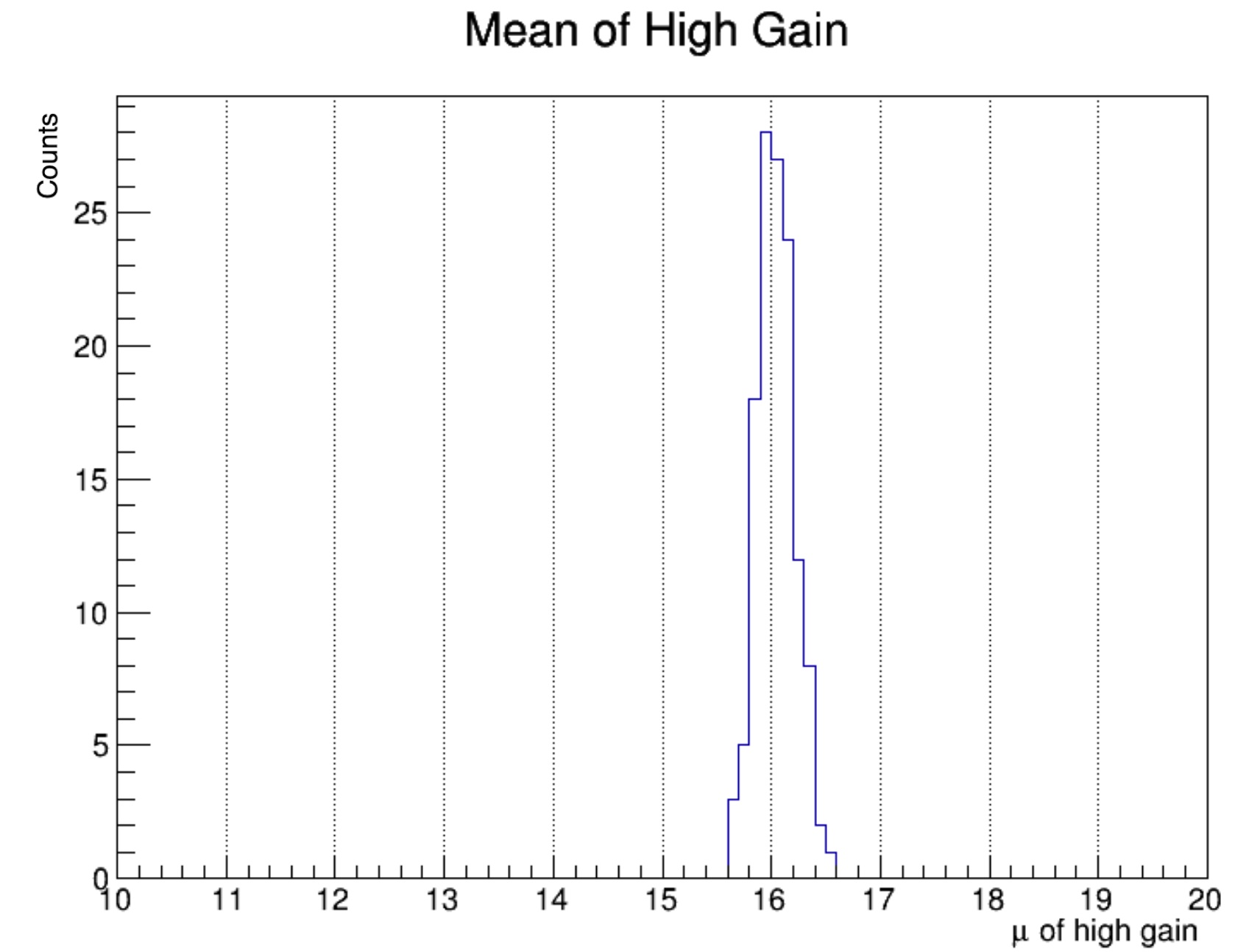}
    \includegraphics[width=0.48\textwidth]{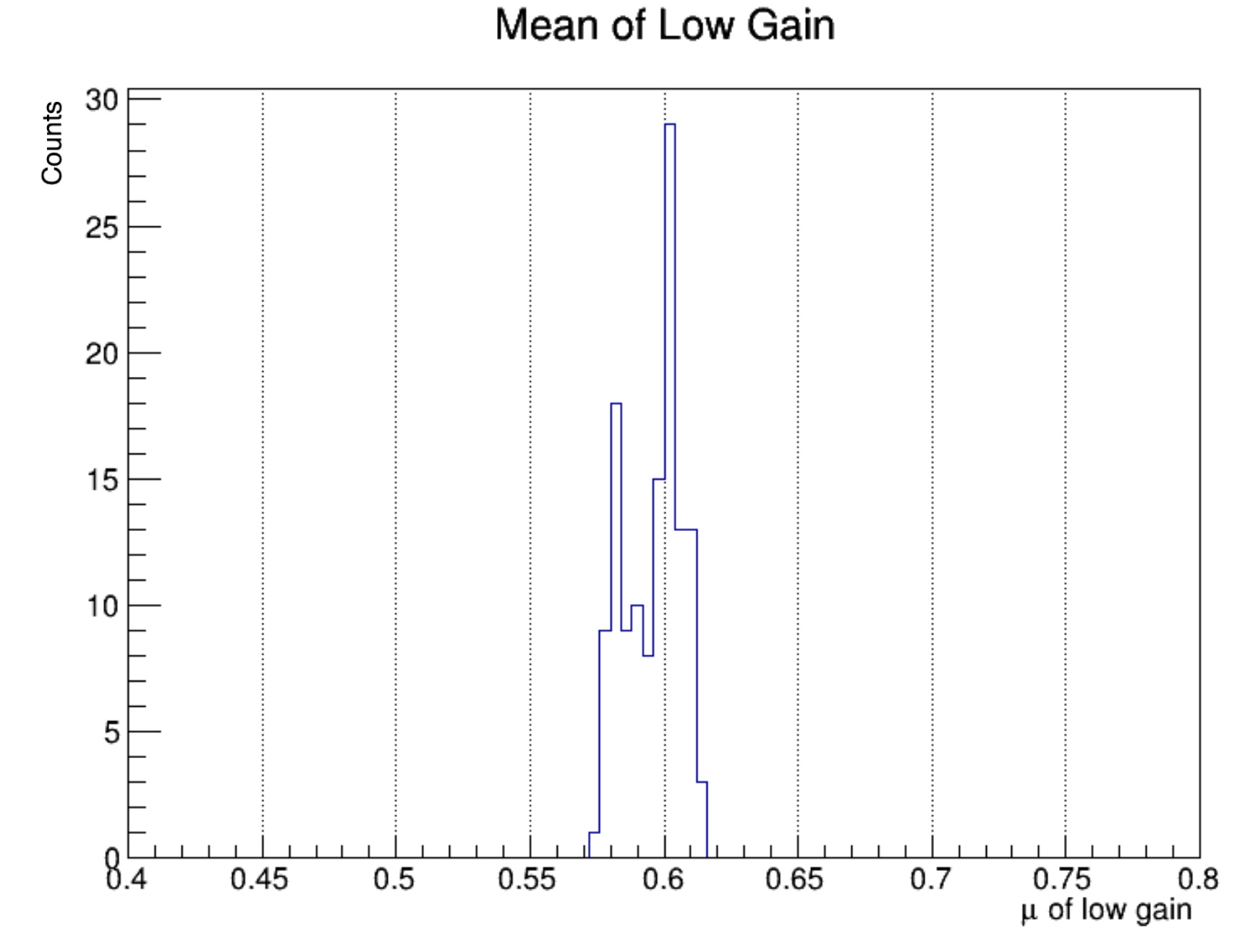}
    \caption{Distribution of gain values across all FEE channels, both high (left) and low (right) gain.}
    \label{fee_gain_distributions}
\end{figure}

The gains of the FEE's high and low gain signal paths are measured to
have average values of 16.0 and 0.59, respectively.
The gain of each FEE channel used in the detector is measured individually, and determined using both ex-situ experimental setups and
in-situ measurements.
Figure~\ref{fee_gain_distributions} shows the distributions of measured gain values for all channels in
the detector.


\section{Trigger and Data Acquisition}

The JSNS$^2$ experiment aims to collect physics signals coming from the collision of the 3 GeV proton beam accelerated by the J-PARC RCS with a mercury target.
The Radio-Frequency (RF) module of the accelerator provides a timing signal synchronized with extraction of the proton beam from the RCS.
The timing signal is useful to detect the arrival of the beam since it arrives at JSNS$^2$ 100\,$\mu$s before the real beam.

\subsection{Trigger Scheme}
Three different types of trigger are utilized to collect physics signals from the beam collision with the mercury target and for detector calibration. The first type is a 25~Hz periodical trigger produced from the RF module and synchronized with the beam collision, named the ``kicker trigger''. A ``beam-off trigger'' is introduced by copying the kicker trigger with a timing delay. The delay is occasionally changed from 11~$\mu$s to 30~ms, to monitor events in the beam-off period.
The second type is a detector ``self-trigger'' formed from the analog sum of the inner PMT signals. The third type is a ``LED trigger'' implemented for the LED calibration described in Section 9.

\begin{figure}[!h]
\centering\includegraphics[width=1.0\linewidth]{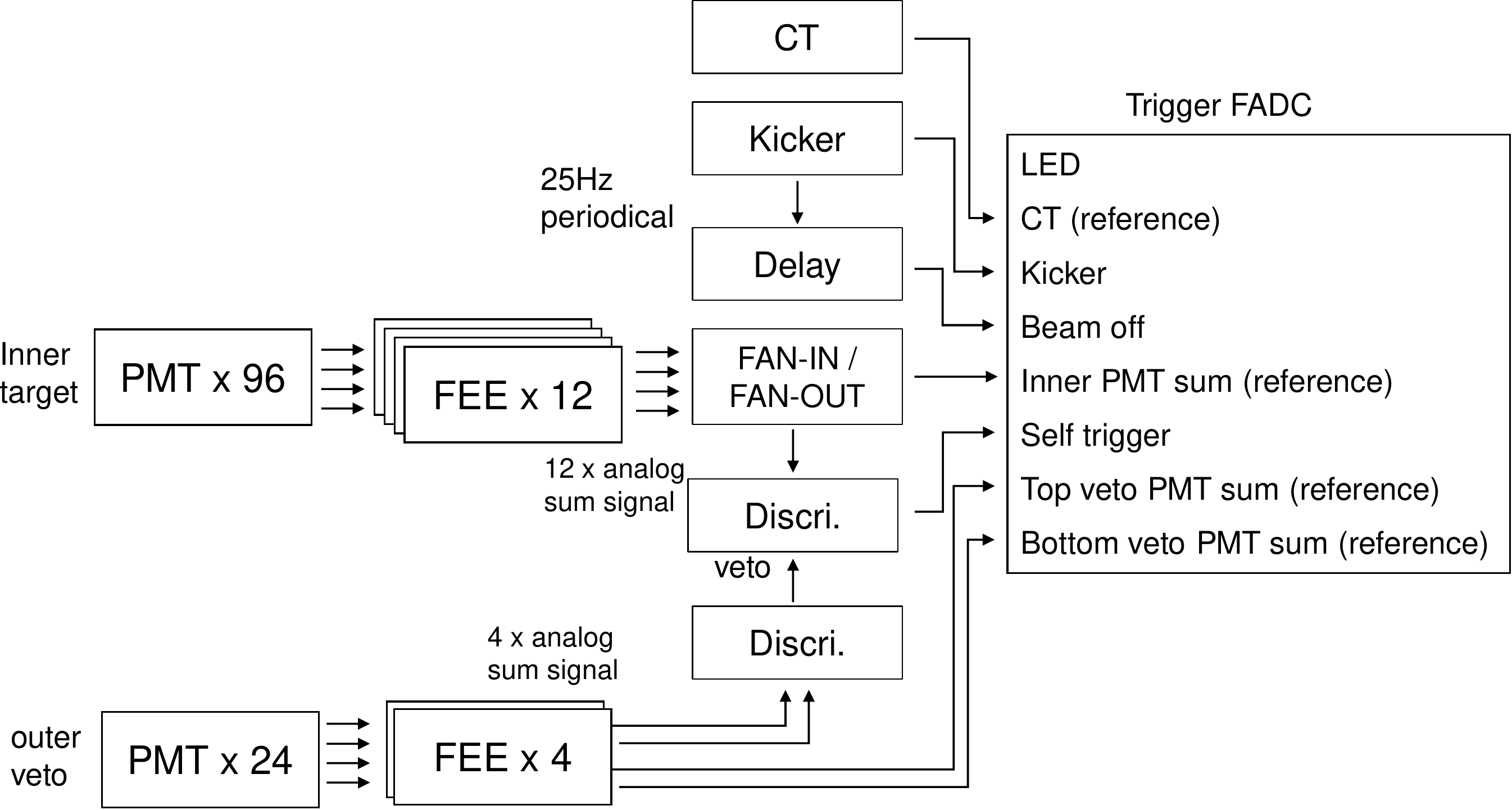}
\caption{A block diagram of the trigger scheme. All boxes except for the PMTs indicate the hardware electronics. The arrows show the paths of analog or NIM signals. The inner and veto PMTs are summed by the FEE and the FAN-IN/the FAN-OUT modules to issue NIM signals by the discriminator (Discri.\ in the diagram). The NIM signals of the  veto PMTs are connected to the discriminator, for the inner PMTs to veto the self trigger. The kicker and CT signals are 25~Hz periodical NIM signals connected to the trigger FADC board. The beam off trigger is created as a copy of the kicker signal with flexible delay time. The CT and analog signals of the inner and veto PMTs are not used for trigger, but these waveforms are recorded for reference.}\label{fig:trigger_scheme}
\end{figure}

One of the FADC boards is used for making a trigger decision, as the CAEN V1721 is able to issue a NIM signal to the other FADC boards using its own internal discriminator. We call this board the ``trigger FADC'' board. The hardware trigger logic is summarized in Fig.~\ref{fig:trigger_scheme}.
The waveforms of the trigger signals are recorded by the trigger FADC board. The waveforms of the analog sum and NIM signals are also recorded for offline analysis. The kicker signal is always produced by the RCS, regardless of whether the proton beam really arrives at the mercury target or not. In order to determine if the proton beam is injected from the RCS accelerator to MLF, we use the reference signals from the current transformers (CTs) located at the beam line.
The CT signal is discriminated and transported to the experimental groups in MLF. The CT signal is approximately 2.7~$\mu$s delayed with respect to the real beam~\cite{CT}.

The analog sum of the veto PMTs is used to form a 100~ns veto-gate signal, called ``online veto'', to suppress the trigger rate due to cosmic rays during the self-trigger data taking. The online veto is also applied during dedicated calibration data taking runs. The veto analog sums are also recorded and read out with each event.

A special trigger scheme, consisting of two successive trigger signals, is implemented for the sterile neutrino search, in order to suppress the background rate and maximize the IBD candidate rate. A prompt trigger is formed if the target PMT analog sum is above 250\,mV, which is 100\% efficient at 15\,MeV, in a roughly 10\,$\mu$s timing window issued at 1.1\,$\mu$s after the first beam bunch. A delayed trigger is formed after a prompt trigger if an analog sum is found above 100 mV, which is 100\% efficient at 5\,MeV, in a subsequent $\sim$12\,ms timing window. The LED and beam-off trigger channels shown in Fig.~\ref{fig:trigger_scheme} are used for the prompt and delayed triggers during the normal data-taking with the special trigger.
The special trigger is optimized to take a maximum number of IBD candidate events by adjusting the energy threshold of the prompt trigger and the timing width of the delayed trigger.

\subsection{Data Acquisition}

\begin{figure}[!h]
\centering\includegraphics[width=1.0\linewidth]{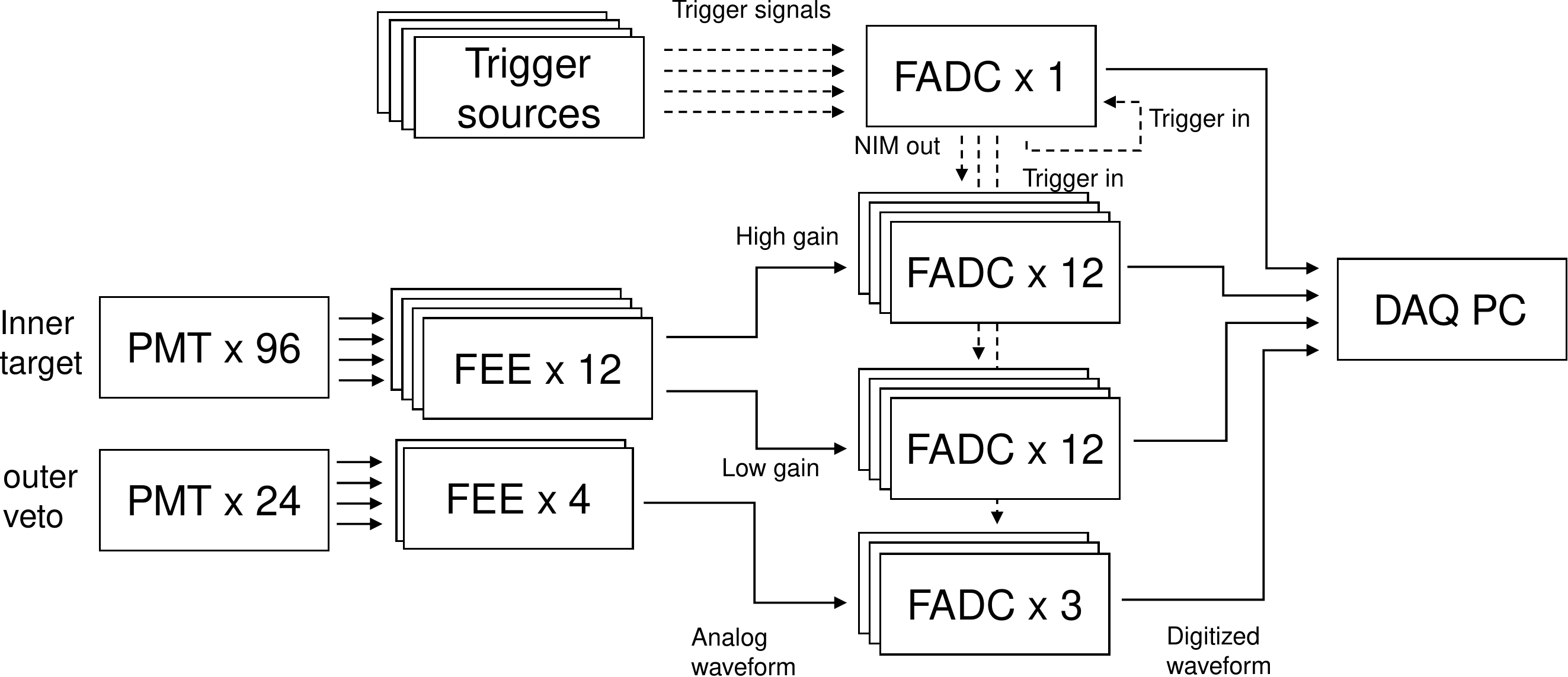}
\caption{A diagram of the DAQ system. The boxes indicate hardware electronics and PC, while the solid and the dashed arrows show analog or digitized waveforms from the PMTs and the trigger NIM signals, respectively. Each of the FEE connected to an inner PMT has two outputs with the higher and lower gain amplitudes, while one FEE of a veto PMT has only a single output. The waveforms from the PMTs and the trigger sources are finally digitized and transferred to the DAQ PC. The NIM out of the trigger FADC is distributed to the FADC boards, including the trigger FADC itself.}
\label{fig:DAQ_scheme}
\end{figure}

All waveforms are collected by four daisy chains of the optical links among the FADC boards. Each of the FADC boards provides meta information about each trigger, such as a 32-bit event counter, and a trigger timestamp. We use the event counter and timestamp to validate the trigger synchronization across the FADC boards. Figure~\ref{fig:DAQ_scheme} shows an overall diagram of the data acquisition (DAQ) system.

The clocking system for the FADC boards also uses a daisy chain, each FADC's clock output is connected to the clock input of the neighboring FADC board.
Therefore, there is a fixed time difference due to the clock delay between neighboring boards.
The relative delay for each PMT, including the board-to-board delay, is measured using LED calibration data and corrected for in offline analysis.
The fixed board-to-board timing offsets are corrected for using the FADC's internal delay functionality.

The FADC board has the capability to change its waveform readout time window size. For the kicker and beam-off trigger data taking, a 10\,$\mu$s time window, equivalent to 5000 FADC samples per channel, is used, followed by another trigger with a 10\,$\mu$s window. A 0.5\,$\mu$s, 250 sample configuration is used for self-trigger data taking. The maximum sustainable self-trigger rate for our readout system is 550 Hz. A 2\,$\mu$s, 1000 sample window is used for the special trigger.

The threshold settings used for typical self-trigger data taking runs can produce trigger rates beyond the limitation of the data bandwidth from the FADCs to the readout PC.
When this occurs, and the FADC memory buffers are full, we pause the trigger and read out the data until the FADC's memory buffers are empty, then resume normal data taking.

\subsection {Data Acquisition Control and Monitoring}

The JSNS$^2$ run control software is based on the EPICS (Experimental Physics and Industrial Control System) framework. Both the readout and DQM (Data Quality Monitoring) PCs have EPICS client programs to launch their DAQ process.
A GUI, built using Control System Studio, is used to start, stop, and reconfigure each DAQ processes.
Run information, such as run start and stop time, and run type, are recorded into a MySQL database running on the DQM PC. File information, such as timestamp and data size, are also recorded into the database. The trigger rate and event size are also recorded into the database. These slow monitoring data are visualized on a custom webpage, built using the Grafana framework~\cite{Grafana}.


\section{Slow Control and Monitoring System}

A slow control and monitoring system (SCMS) is implemented to provide reliable control of detector operation and quick monitoring of both operational status and environmental conditions.
The SCMS issues an alarm if any of its monitored the values exceed a preset acceptable range.
The system includes monitoring of the liquid level in the detector and rapid alert of any possible liquid overflow and leakage.
It also monitors the temperature and air pressure inside the detector, the humidity in the experimental area, and the liquid flow rate during filling and extraction.
The SCMS is also responsible for monitoring and controlling the HV supplied to each PMT.

\begin{figure}[h]
\begin{center}
\includegraphics[scale=0.32]{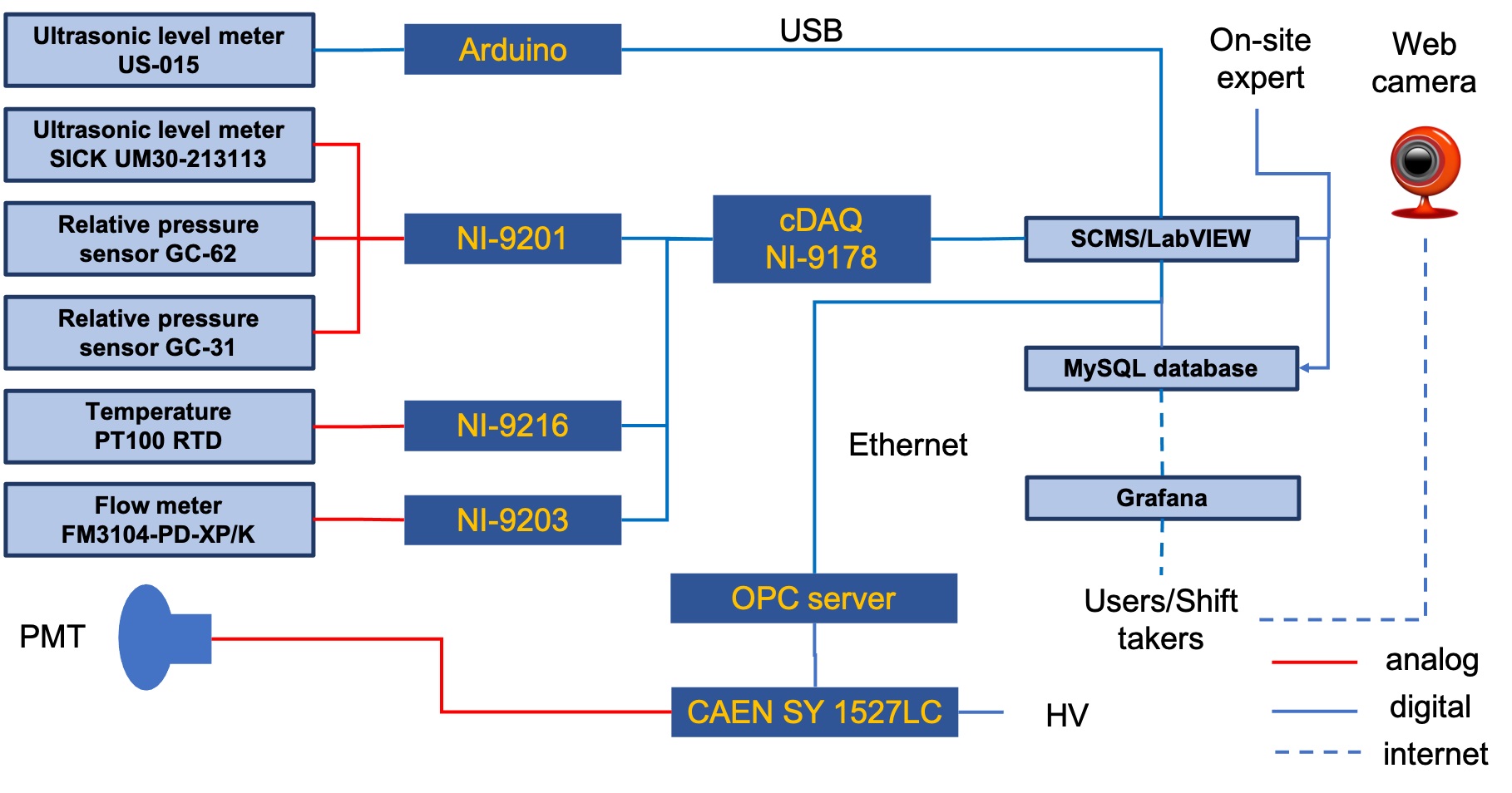}
\end{center}
\caption{\setlength{\baselineskip}{4mm}Conceptual diagram of the JSNS$^2$ slow control and monitoring system.}
\label{fig:schematic}
\end{figure}

  The data acquired by sensors are collected by readout modules and forwarded to a LabVIEW-based client program~\cite{LabVIEW} for monitoring and storing. The monitored data are visualized for displaying detector operational status using a Grafana based graphical user-interface tool. The data are also stored in a MySQL database~\cite{MySQL} every 30 seconds.
Figure~\ref{fig:schematic} shows a block diagram of the JSNS$^2$ SCMS to show its overall configuration.

The LabVIEW program queries sensors every 5 seconds and displays them over an 8-hour timespan. The LabVIEW display includes liquid levels, liquid leak in the spill tank, detector temperature, detector pressure, and humidity in the detector area.

\begin{figure}[h!]
\begin{center}
\includegraphics[scale=0.5]{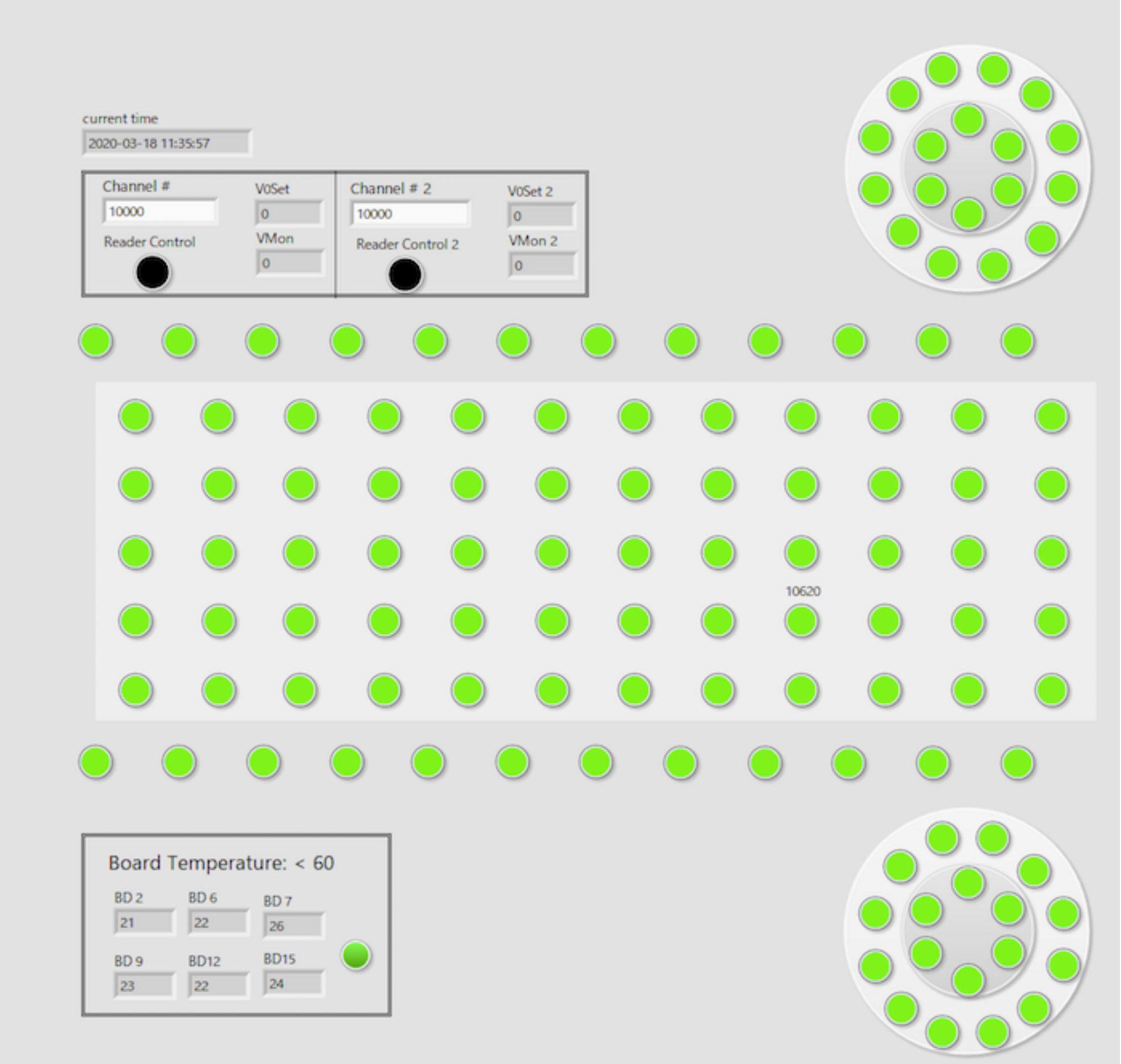}
\end{center}
\caption{\setlength{\baselineskip}{4mm}A screenshot of HV control and monitor display. The green circle indicates that its PMT HV is within 10\,V range of its preset value. The temperature of a HV module is also displayed.}
\label{HV_LabVIEW}
\end{figure}

A LabVIEW-based HV control and monitoring program integrated into the SCMS is responsible for controlling HV to each PMT, storing preset HV values, and reading applied HV values and the temperatures of the HV modules. The CAEN operational process control (OPC) server is used to control and communicate with the HV modules~\cite{OPC}. The LabVIEW program obtains data from the OPC server via Ethernet and displays the status of individual HV supplied to each PMT in an exploded view of the detector. Figure~\ref{HV_LabVIEW} shows a screenshot of the PMT HV display where the color represents the difference between the preset and currently supplied HV values.

The Gd-LS and LS levels are measured with SICK ultrasonic sensors~\cite{SICK-sensors}.
A long-range UM30-215113 sensor is used for liquid filling and extraction and a short-range UM30-213113 sensor is used during normal data-taking.
The measured level generates an analog voltage value, 0 to 10\,V, which is read out by a National Instruments (NI) 9201 module~\cite{NI9201}.
Four US-015~\cite{US015} ultrasonic sensors are used to monitor for a possible liquid leak inside the spill tank, and two additional US-015 sensors are used to monitor the liquid level inside the stabilization container;
these sensors are read out by an Arduino Uno Rev3~\cite{Arduino}.

The detector is sealed to avoid oxygen contamination into the LS and Gd-LS.
This allows a pressure difference between the inside and outside of the detector to develop. For monitoring this pressure difference, two types of relative pressure meters are installed. The GC-31~\cite{GC-31} and GC-62~\cite{GC-62} both provide voltage outputs proportional to the pressure difference. These voltages are read out by an NI 9201 module.

A total of eight PT100 RTD sensors measure temperatures in the detector veto region. The temperature outputs are read out by an NI 9216 module~\cite{NI9216}.

The liquid flow rate during filling into or extraction from the detector is measured by a flow meter.
The flow meter provides an analog current proportional to the flow rate. An NI 9203 module~\cite{NI9203} reads and digitizes the current output.

An NI cDAQ-9178 crate~\cite{NI-crate} houses our three NI readout modules, and is connected to the SCMS PC via a USB cable. The SCMS PC runs a LabVIEW readout program to receive and record sensor values. LabVIEW was selected because it interfaces well with the NI modules.

A TR-73U ambient sensor~\cite{TR-73U} monitors environmental conditions around the detector such as temperature, humidity, and atmospheric pressure. Its recorded data are read out and delivered to the LabVIEW program running on the SCM PC via USB.

The SCMS, as described here, has been successfully tested, commissioned, and deployed.
Using this system liquid filling and extraction for the JSNS$^2$ detector has been completed
without significant trouble. In addition to its continued use to support reliable
data-taking, this demonstrates that the SCMS provides sufficient monitoring and safety assurance
for the experiment. A more detailed description of the SCMS can be found in Ref.~\cite{PTEP-SCMS}.


\section{Data Processing}

\begin{figure}[h]
\centering\includegraphics[width=1.0\linewidth]{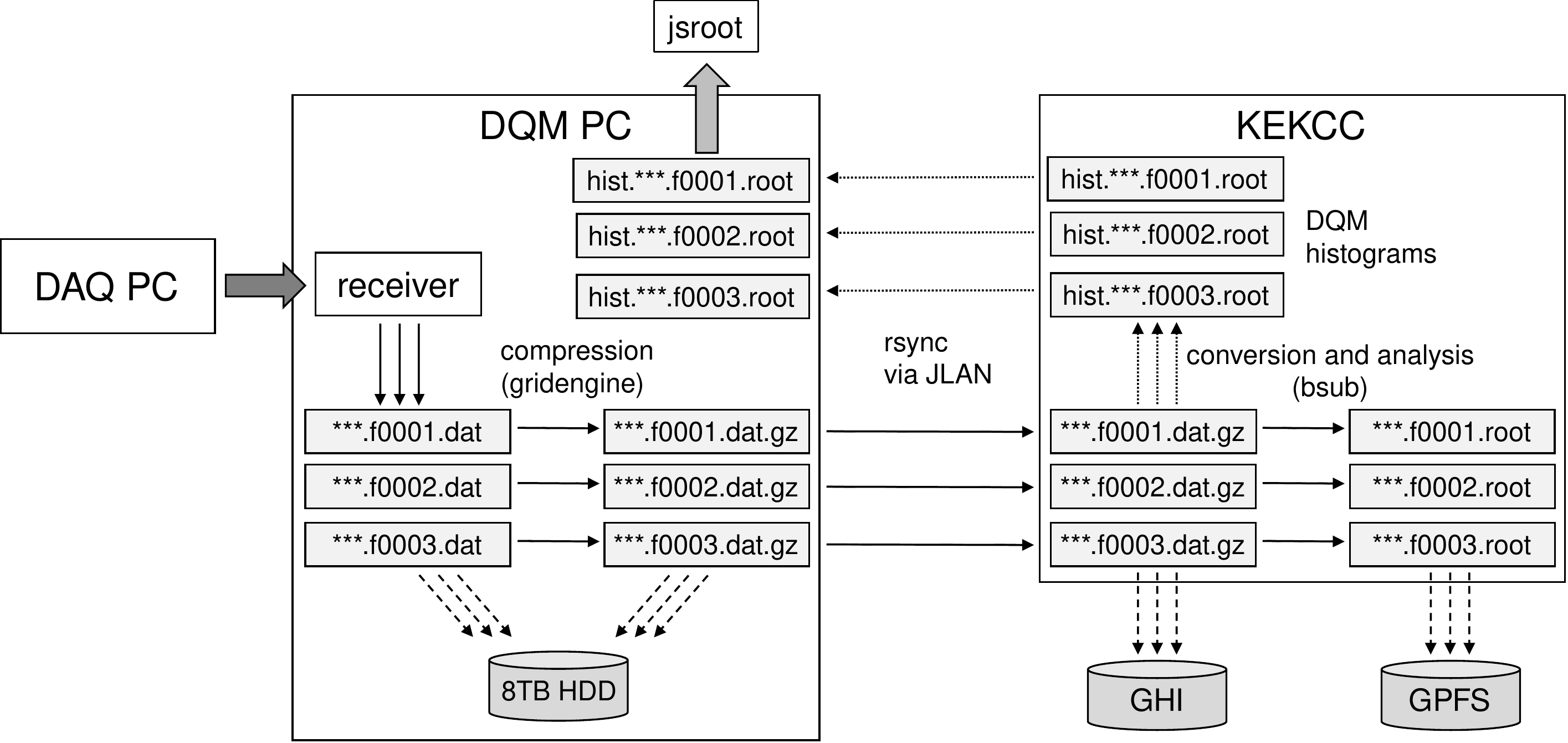}
\caption{A schematic of the data processing scheme. The boxes show PCs and files. The solid lines show the data transfer between processes over the network, while the dashed lines show the data paths to the storage systems.}
\label{fig:data_processing}
\end{figure}

The JSNS$^2$ data processing system is responsible for transporting the waveform data from the DAQ system to the KEK Computing Center (KEKCC), and for performing a quick data analysis routine as part of the DQM system.
The data processing system consists of two parts:
(1) processing at the on-site PCs as preparation for data transfer and (2) processing at the KEKCC by performing file type conversion and a quick analysis to save into the disk.
Figure~\ref{fig:data_processing} shows the data processing scheme.

For the initial part of data processing the DQM PC receives data from the FADC boards via 1\,Gbps Ethernet and dumps them into binary files in 8\,TB RAID-5 disk as temporary storage, sufficient for a week of data taking. A parallel job system is used on the DQM PC to compress each binary files. The compression factor is about 3 to 6. The files are finally transferred to KEKCC via rsync with the SSH protocol; the DQM PC is connected to KEKCC by a 1\,Gbps Ethernet connection using J-PARC's internal LAN.

As the second part of the data processing, the compressed files are posted at KEKCC's GPFS HDD based file system. The DQM PC submits jobs to KEKCC to convert the data files into ROOT. The produced ROOT files contain ROOT TTree objects and are used for offline analysis. The conversion is done in a few minutes on average. Finally, the compressed files are moved to the tape drive system in KEKCC for long term storage, while the ROOT files are on the GPFS for user analysis.

\subsection {Data Quality Monitor}
In parallel to the file transfer, the online data and the recorded files are analyzed at the on-site DQM PC and at KEKCC, respectively, to extract several detector status values. Since the CPU power in the DQM PC is limited, a brief analysis routine is integrated with the online data stream on the DQM PC to produce information, such as raw waveforms of PMTs. A further analysis is done in KEKCC using the converted root files, which extracts more information. Basic information such as trigger bit rate is obtained in an analysis to monitor the beam injection consistency. The results from both the on-site PC and KEKCC are formatted as histograms in ROOT files and displayed on a webpage. The webpage uses JSROOT to read the histogram files.

\subsection {Data Production}
The waveforms in the data files are quickly analyzed to extract signal pulses as charge and timing information for each PMT. We developed a common framework for data analysis named JADE (JSNS$^2$ Analysis Development Environment), which reconstructs PMT pulses and identifies events in the long waveforms. For higher level analysis, JADE is also used to do vertex position and energy reconstruction. Data are processed in parallel using JADE and KEKCC’s batch job system.


\section{Calibration System}

Three sources are used to calibrate the response of the detector. An LED system is used to calibrate the relative timing of the individual PMTs as well as the gain. A neutron source is used to calibrate the energy scale at several positions in the detector, with the 8~MeV energy peak coming from neutron capture on Gd. Finally, the Michel electron spectrum from stopping cosmic muons is used to calibrate the energy response and the energy resolution up to the 52.8~MeV end-point.

In this section, the three kinds of calibration are described in more detail.

\subsection{LED system}

\begin{figure}[!htbp]
  \centering
    \includegraphics[width=0.48\textwidth]{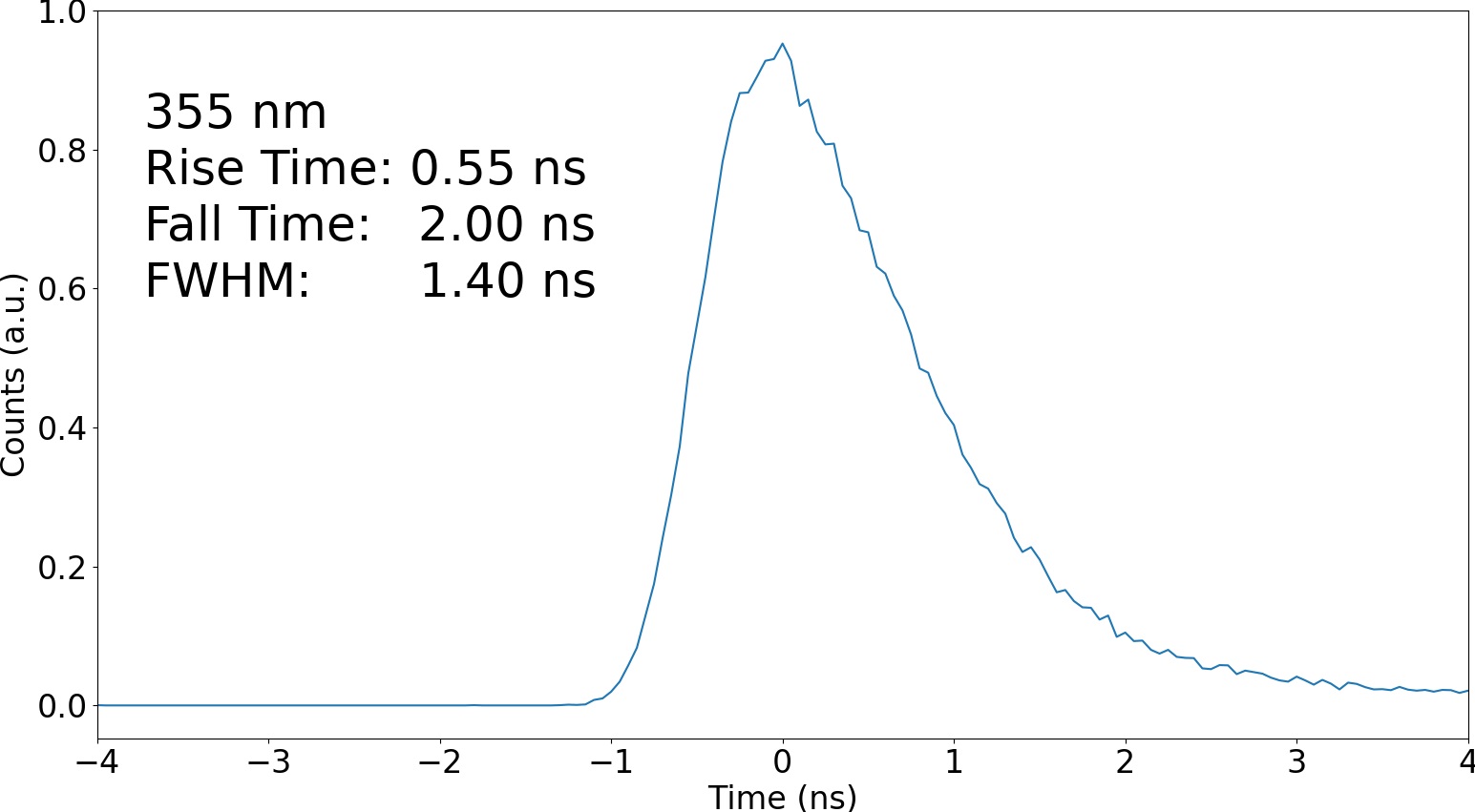}
    \includegraphics[width=0.48\textwidth]{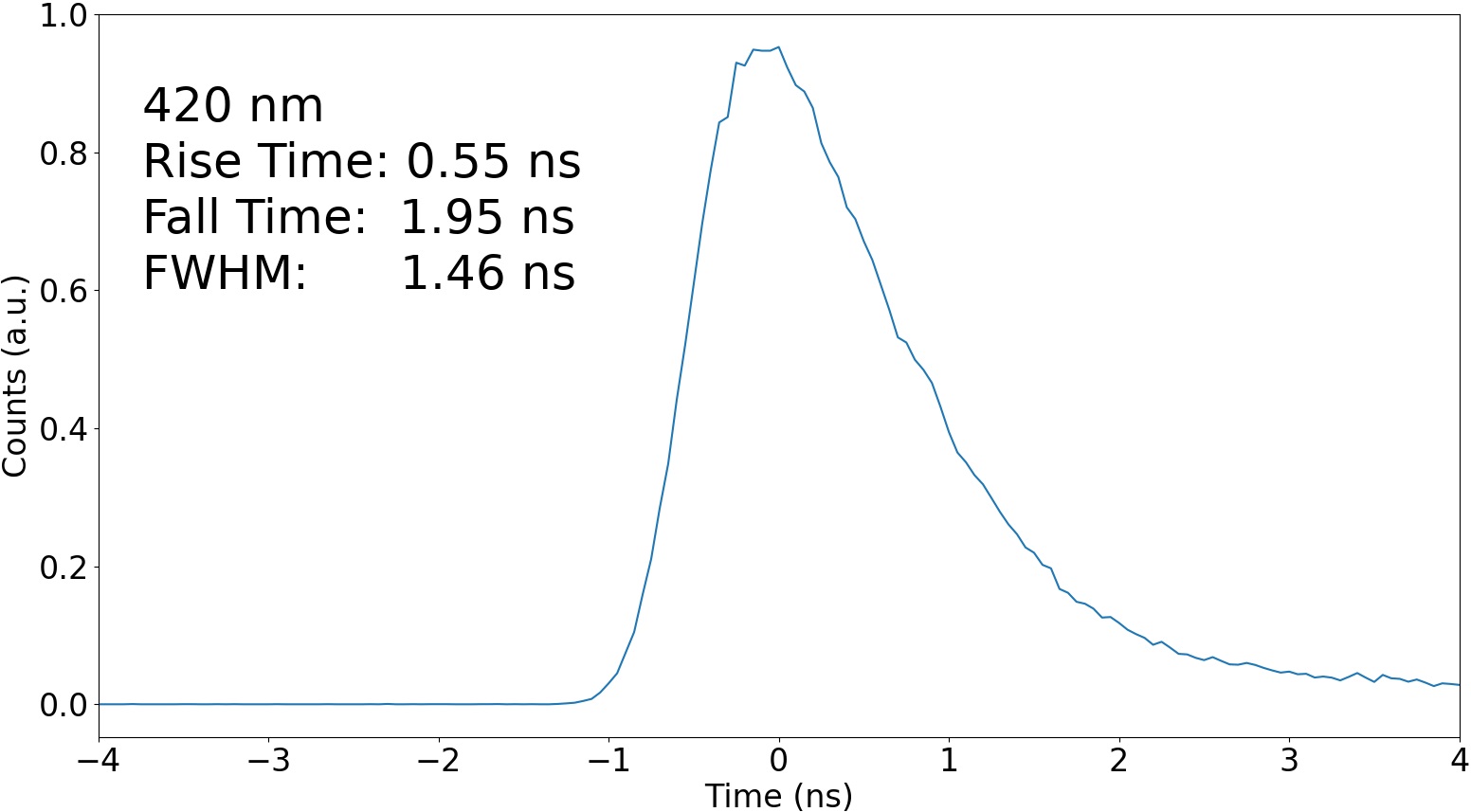}
    \caption{Pulse profiles of the 355 (left) and 420\,nm (right) LEDs.}
    \label{fig:LED-pulse}
\end{figure}

The JSNS$^2$ LED system is based on the LED systems that have been used in Double Chooz and SNO+~\cite{snoplus_led}. For those systems, where ensuring low-radioactivity backgrounds is a key requirement, the electronics were installed outside the detector, and light pulses enter the detector via optical fibers.  As accurate timing is a key requirement for JSNS$^2$ and the energy depositions from the physics signals are above the natural radioactivity backgrounds, the LEDs in this system are embedded in a sealed acrylic housing and mounted on the PMT array. This enables an LED optical pulse to enter the detector without any degradation to the pulse width. Lab tests have demonstrated that the fast signals in the electronics inside the detector do not produce any interference with the PMT signals. In addition, the Kapustinsky-based LED drivers~\cite{kapustinsky} used by SNO+ have been much improved for this system~\cite{led_patent}: The optical pulses have a rise time of 0.55~ns, as shown in Fig.~\ref{fig:LED-pulse}. These measurements have been directly obtained using the method described in~\cite{timing_meas}.

\begin{figure}[!htbp]
  \centering
  \includegraphics[width=0.70\textwidth]{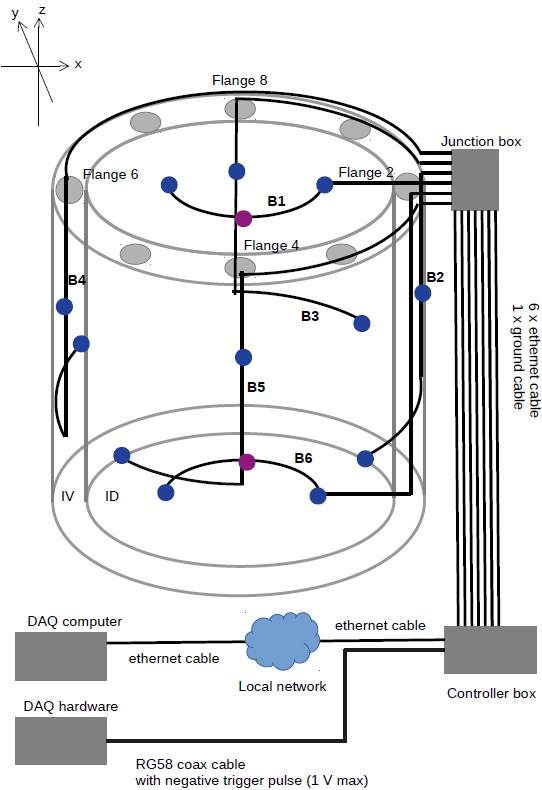}
  \includegraphics[width=0.19\textwidth]{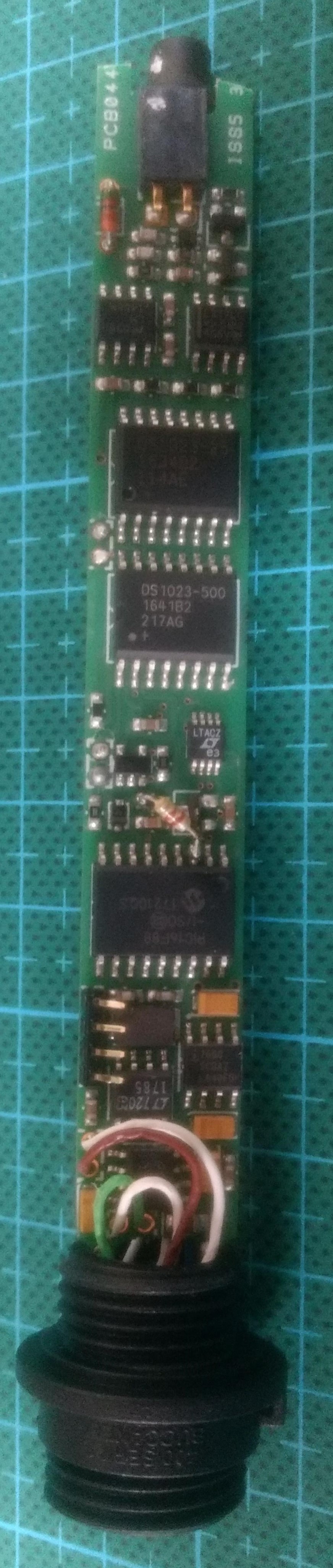}
  \caption{A schematic overview of the JSNS$^2$ LED system (left). The (two) purple points indicate the positions of the 355~nm LEDs, the (twelve) blue points the positions of the 420~nm LEDs. All LEDs point inwards. The numbers B1--B6 indicated the different branches, which connect two or three LEDs. IV represents the Inner Veto volume and ID represents the Inner Detector volume. The LED driver circuit (without the acrylic housing) is shown on the right, with the flattened LED at the top.}
  \label{fig:led_system_overview}
\end{figure}

The JSNS$^2$ LED system consists of twelve 420~nm LEDs and two 355~nm LEDs. An overview of the system, as well as a photo of the driver circuit of the LEDs inside the detector, is shown in Fig.~\ref{fig:led_system_overview}.

Standard 5~mm LEDs (T-1 3/4 package)~\cite{UV-LED} are used. To ensure the maximum opening angle, the curved front is removed and the remaining flat end is polished. This enables full coverage of the detector for the 420~nm LEDs, for which the scintillator is effectively transparent. The light of two 355~nm LEDs gets absorbed and re-emitted uniformly inside the target. LED drivers can be individually addressed. A selected driver returns a trigger pulse in-sync with the LED pulse to the controller box. After calibration, the controller box provides central data-acquisition a trigger signal in-sync with the LED pulse. The timing error between the LEDs and jitter is significantly less than the rise time of the optical pulse. The intensity can be controlled and ranges from less than 50 to over $10^6$~photons per pulse.

\subsection{Neutron source}

\begin{figure}[!htbp]
  \centering
  \includegraphics[width=0.60\textwidth]{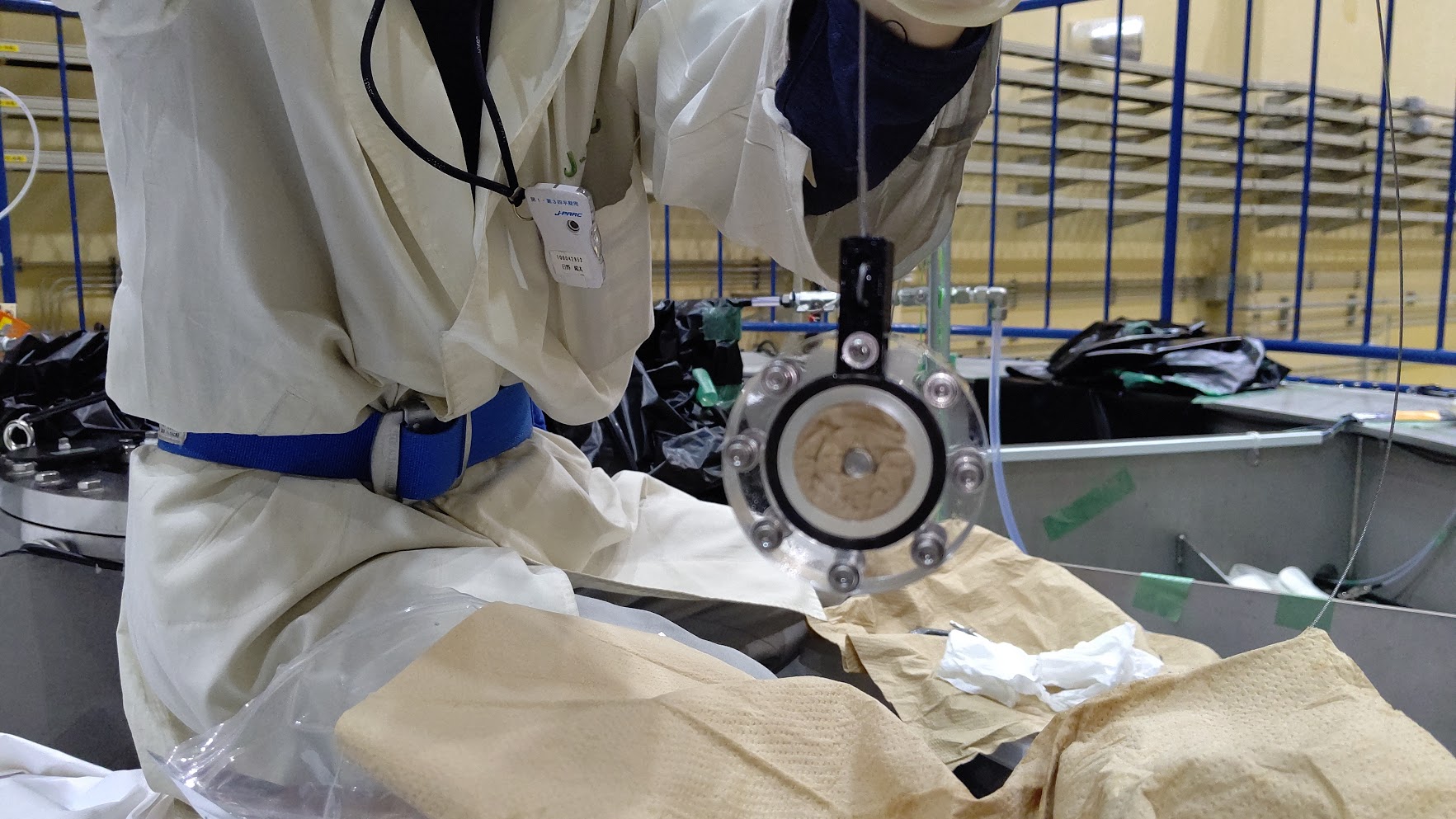}
  \caption{The JSNS$^2$ $^{252}$Cf calibration source inside an acrylic container}
  \label{fig:cf_source}
\end{figure}

The neutron calibration source is used for energy calibration, in particular to convert the observed number of photoelectrons to energy, at several positions along the central axis of the detector.
Neutrons are produced by a $^{252}$Cf radioactive source with outer dimensions of 20~mm in length and 5~mm in diameter, and with a calibrated activity of $3.589\times10^6$~Bq on the 10$^{\rm th}$ of August 1983.  The dominant capture is expected to be on $^{157}$Gd having 15.65\% natural abundance and a capture cross-section of 254,000~b, resulting in 7.937~MeV total energy in gammas. A weaker second peak is expected to be observed at 8.536 MeV from the capture on $^{155}$Gd having 14.80\% natural abundance and a capture cross-section of 60,900~b~\cite{n-Gd-gamma}.
A short run of several minutes provides sufficient events for a precise calibration.

\begin{figure}[!htbp]
  \centering
  \includegraphics[width=1.0\textwidth]{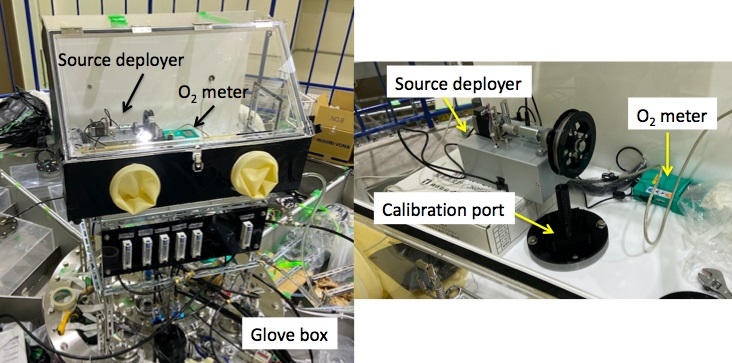}
  \caption{Glove box for deploying a neutron source. It is filled with nitrogen gas and deploys a $^{252}$Cf source inside an acrylic container using a stepping motor controlled by a PC.}
  \label{fig:glove-box}
\end{figure}

The neutron source is contained in an acrylic and cylindrical holder, as shown in Fig.~\ref{fig:cf_source}, with dimensions of 60~mm in diameter and 16~mm in thickness. The neutron source, in the acrylic container, can be deployed at any detector position along the central axis. A source deploying device is made of a disc-shaped reel with a diameter of 100~mm, a stainless steel wire, and a stepping motor as shown in Fig.~\ref{fig:glove-box}. The source can be deployed at a desired location by rotating the reel using the stepping motor and adjusting the wire length accordingly. The motor is controlled by a micro-controller using a PC. In order to avoid oxygen and moisture contamination, a glove box filled with nitrogen gas is installed as a calibration port above the acrylic chimney. Figure~\ref{fig:glove-box} shows the glove box where the neutron source is deployed into the detector.

\subsection{Michel electrons}

Michel electrons coming from cosmic muons stopping in the JSNS$^2$ detector provide a copious event sample for the detector calibration. They are useful for understanding the detector response in the entire energy range of the IBD prompt signal. Because of their uniform event vertices, they can be used for surveying a position dependence of the detector response.
During a recent commissioning run, $\sim100$~Hz of stopping muons were observed inside the central volume with a rate of $\sim2400$~Hz cosmic muons.
The high statistics of Michel electrons allow monitoring the stability of the detector response over time as well as a possible in-situ measurement of the optical attenuation length of Gd-LS.

\section{Summary}
\noindent

The JSNS$^2$ detector is designed and built to be sensitive to $\overline{\nu}_e$ of energies less than 52.8 MeV, motivated by the sterile neutrino search.
The construction of the JSNS$^2$ detector was completed in the spring of 2020.
Following the completion of the detector the experiment was ready for a search for sterile neutrinos using the neutrino beam from muon decay at rest.
During June 2020 the first 10 days of physics data with the neutrino beam was acquired. After data taking, the Gd-LS and LS liquids were extracted back to the ISO tanks, and the detector was relocated and stored at the detector storage building for a maintenance period, until December 2020.
For the improved PSD performance, an additional solvent DIN was dissolved into Gd-LS before the second run began in January 2021. The PMT gain and electronic dynamic range are adjusted for efficient IBD event detection.
For reliable and accurate event reconstruction, the detector is calibrated using an LED system, a Cf neutron source, and Michel electrons produced by cosmic muons.
A slow control and monitoring system was implemented to provide reliable control of detector operation and quick monitoring of operational status and environmental conditions.
Based on the first and second runs, we have evaluated the detector operation including triggering, data acquisition, and data processing. Necessary modifications and improvements have been performed to carry out a successful search for sterile neutrino oscillations.

\section{Acknowledgement}
We thank the J-PARC staff for their support. We acknowledge the support of
the Ministry of Education, Culture, Sports, Science, and Technology (MEXT) and
the JSPS grants-in-aid (Grant Numbers 16H06344, 16H03967, and 20H05624), Japan.
This work is also supported by the National Research Foundation of Korea
(NRF) Grant No. 2016R1A5A1004684, 2017K1A3A7A09015973, 2017K1A3A7A09016426,
2019R1A2C3004955, 2016R1D1A3B02010606, 2017R1\\A2B4011200, 2018R1D1A1B07050425,
2020K1A3A7A09080133 and 2020K1A3\\A7A09080114. Our work has also been supported by
a fund from the BK21 of the NRF. The University of Michigan gratefully acknowledges the
support of the Heising-Simons Foundation. This work conducted at Brookhaven National
Laboratory was supported by the U.S. Department of Energy under Contract DE-AC02-
98CH10886. The work of the University of Sussex is supported by the Royal Society grant
no.IESnR3n170385. We also thank the Daya Bay Collaboration for providing the Gd-LS liquid,
the RENO collaboration for providing the LS liquid and PMTs, CIEMAT for providing the splitters,
Drexel University for providing the FEE circuits, and Tokyo Inst. Tech for providing FADC boards.


\bibliography{mybibfile}

\end{document}